\documentclass[prd,a4paper,eqsecnum,nofootinbib,preprint,floatfix,showpacs]{revtex4} 
\usepackage{epsfig} 
\usepackage{colordvi} 
\usepackage{graphicx} 
\usepackage{bm}

\def\tr{\text{tr}\,} 
\def\Eq#1{Eq.~(\ref{#1})}

\def\<{\langle} 
\def\>{\rangle}
 
\newcommand{\text}{\rm} 
\def\>{{\rangle}} 
\def\<{{\langle}}

\def\tr{{\text tr}\,} 
 
\def\Eq#1{Eq.~(\ref{#1})} 
\def\e{{\text e}}

\begin{document} 
 
\vspace*{0.5in} 
 
\title{Chiral crystals in 
strong-coupling lattice QCD\\
 at nonzero chemical potential\vspace*{0.25in}} 
\author{Barak Bringoltz\\ 
\vspace*{0.2in}} 
 
\affiliation{Rudolf Peierls Centre for Theoretical Physics, 
University  of  Oxford,\\ 
1 Keble Road, Oxford, OX1 3NP, UK \\ 
\vspace*{0.3in}}

\begin{abstract} 
We study the effective action
for strong-coupling lattice
QCD with one-component staggered fermions
 in the case of nonzero chemical
potential and zero temperature. The 
structure of this action
suggests that at large chemical potentials
its ground state is a crystalline `chiral 
density wave' that spontaneously breaks chiral symmetry and
translation invariance. In mean-field
theory, on the other hand, we find that this state is unstable. We show that lattice artifacts are
partly responsible for this, 
and suggest that if this phase exists in QCD, then finding it in
  Monte-Carlo simulations would require simulating on relatively 
fine lattices. In particular, 
the baryon mass in lattice units, $m_B$, should be considerably smaller than its 
strong-coupling limit of $m_B\sim 3$. 

\end{abstract} 
 
\maketitle 
 
\section{Introduction} 
\label{sec:intro} 
 
The idea that the ground state of QCD at
 zero temperature and 
 nonzero chemical potential can
 spontaneously break translation invariance
was first suggested by Deryagin, Grigoriev and Rubakov (DRG)
\cite{DGR}. They studied large-$N$ QCD at weak
coupling, and showed that the four-quark 
interaction, that is generated by a single gluon 
exchange, causes an instability
in the quark Fermi sea toward the creation of a condensate of the form 
\begin{equation}
\<\bar \psi(x) \psi(x)\>\sim \cos{\left(2\vec Q \cdot \vec x\right)}.
\label{eq:DGR}
\end{equation}
Here the wave vector $\vec Q$ has an 
arbitrary direction but its modulus is given by the Fermi
 momentum of the quarks
\begin{equation}
Q = p_F. \label{Q_pF}
\end{equation}
This phenomenon is often referred to as
chiral density waves, and
has famous solid-state counterparts like the 
one-dimensional Peierls instability
 \cite{Peierls}, and the
spin density wave of the Overhauser effect
\cite{Overhauser}.

The large-$N$ result of \cite{DGR} was consequently found to be misleading
\cite{Shuster,Rho}, and Shuster
 and Son showed in \cite{Shuster} that if $N$ is
 decreased from infinity to $N\stackrel{<}{_\sim}O(1000)$,
then color-superconductivity becomes 
 preferable over the DGR state.
 Nonetheless, the authors in
\cite{Shuryak} studied what may happen at lower densities (for $N=3$), where weak
coupling treatments are less reliable
\cite{Rajagopal_Shuster}. The
 conclusion of \cite{Shuryak}
was that chiral density waves may still be competitive
with superconductivity, especially when one uses an instanton
vertex to couple the quarks. 

Recently, there has been related progress in the
Gross-Neveu model. This $1+1$
 model can be studied in large-$N$, and its phase structure 
in the temperature-density plane 
was known to have a line of first 
order transitions that separate a low density phase with
spontaneously broken
 chiral symmetry, and a high density phase, where this
symmetry is intact.
This structure arises when one restricts
to an $x$-independent ansatz for $\< \bar \psi(x) \psi(x)\>$ and was recently
revisited in \cite{Thies1} (and later with a lattice regularisation in \cite{Urs}), where the condensate was allowed to depend on the spatial coordinate of
$x$. The study in \cite{Thies1} 
discovered a new phase at low temperatures and high
densities, which is separated by a second order phase transition from
the low density phase, and where the chiral condensate has a
crystalline structure. 
This crystal structure was also seen in the `t Hooft model \cite{Thies2}, and is reminiscent
of results from older work in the Skyrme model \cite{Verbaarschot}.

In this paper we look for a phase with chiral density waves 
using an effective
action that is derived from 
the strong-coupling expansion of lattice QCD. Apart
from a phenomenological interest
in this model, and in the way it may expose the
phase we are after, we are also motivated by the following, more practical,
reason. Future
Monte-Carlo simulations 
of QCD at low temperatures and large chemical potentials, that will manage to control the sign problem, will presumably
 start by simulating relatively coarse lattices with strong
 couplings. This makes the knowledge on the phase diagram of the 
strong-coupling limit important, and in our context 
it is desirable to know whether one should expect a crystal phase,
and if so, at which values of the chemical potential  $\mu$. 

Indeed, numerous authors have
 analytically investigated the phase diagram of the strong-coupling 
limit (for reviews see
 \cite{Kawamoto_recent,MPL}), but the
possibility to have chiral
density waves was never considered, and 
 these works were always
restricted to homogeneous field configurations. 

Here we relax this constraint
for the first time. To do so, we
 choose to work with one-component
 staggered fermions. In four dimensions, 
these lattice fermions describe four
degenerate continuum Dirac fermions, commonly referred
 to as tastes. The continuum theory has an $SU(4)\times SU(4)$ chiral symmetry,
that explicitly breaks on the
lattice to the following axial taste 
non-singlet $U(1)$ 
\begin{equation}
\psi  \to  \exp \left( i \theta \, \gamma_5 \otimes \xi_5 \right)\, \psi, \qquad
\bar \psi  \to  \bar \psi \, \exp \left( i \theta \,\gamma_5 \otimes \xi_5 \right).\label{U1eps}
\end{equation}
Here $\xi_5$ is a $4\times 4$ traceless matrix that operates in taste
space, and that can be chosen to have the same matrix elements of
$\gamma_5$. 
The ground state that we study in this paper is characterised by the helical condensates
\begin{eqnarray}
\<\bar \psi(x)\, \psi(x)\>&\sim&  \cos \left( 2\vec Q \cdot \vec
  x\right),\label{gs1} \\
\<\bar \psi(x) \left( i \,\gamma_5 \otimes \xi_5 \right) \psi(x)\>&\sim& 
\sin \left( 2\vec Q \cdot \vec x \right),\label{gs2}
\end{eqnarray}
that spontaneously break the chiral symmetry 
of \Eq{U1eps} as well as translation
invariance. Note that Eqs.~(\ref{gs1})--(\ref{gs2}) cannot be
rotated by a global transformation to have $\<\bar \psi(x)
\left(\gamma_5 \otimes \xi_5 \right) \psi(x)\>=0$. This means that if
we regard the continuum limit as a theory of four degenerate light quarks, then our chiral waves break $SU(4)$ flavor. 
As a result,
we are studying
a ground state which is closer to the one 
studied in \cite{Sadzikowski} than to the original DGR state, which
is a flavor singlet.

The reason we choose this particular type of lattice fermions is that, in the strong-coupling limit, they give rise to a
 very simple effective action in terms of hadrons. This action was derived some time ago by Hoek, Kawamoto, and Smit \cite{HKS}, 
and in the next section we re-derive it for self-completeness. After discussing why it is natural to expect chiral 
density waves in the ground state of this action (at nonzero  $\mu$), we formulate in
 Section~\ref{sec:MFT}
 a mean-field theory that is general enough to 
allow for such chiral waves to emerge. The mean-field equations are solved in 
Section~\ref{sec:MFsol}, where we also investigate how the
 mean-field ground state evolves when we increase $\mu$.
We conclude in Section~\ref{sec:summary}
with a few remarks on the implications of this work and on future prospects. Appendix~\ref{sec:baryon_det}
includes technical details related to
the calculation of the mean-field equations and the mean-field free energy.

\section{The effective action} 
\label{sec:Seff} 
 
In this section we follow the work of Hoek, Kawamoto, and Smit
(HKS) \cite{HKS}, and 
re-derive their effective action. Readers who are familiar with
\cite{HKS} can proceed to Section~\ref{Seff_CDW}, where we discuss
the structure of the effective action, and why it is natural to expect 
a chiral density wave in its ground state at nonzero $\mu$.

\subsection{Deriving the Hoek-Kawamoto-Smit action for hadrons}
The starting point 
is the strong-coupling limit of an $SU(N)$ lattice gauge theory with one-component staggered fermions. The action in this case is
\begin{eqnarray} 
S&=&-\frac12 \sum_{n,\nu} \left[\bar{\chi}_{n} \eta_{n\nu} U_{n\nu} 
\chi_{n+\hat \nu} -\bar{\chi}_{n+\hat \nu} \eta^{-1}_{n\nu} U^\dag_{n\nu} 
\chi_{n}\right] + \sum_n \left[ N J_n M_n + \bar{c}_n B_n + \bar{B}_n c_n\right], 
\label{eq:S_LQCD} 
\end{eqnarray} 
where $n=(n_0,n_1,\dots,n_d)$ is an Euclidean lattice index
 and $\nu= 0, 1, 2,\dots, d$ 
is the direction index ($d$ is the 
number of spatial dimensions). The fields $\chi$ and $\bar{\chi}$ are
independent Grassmann
variables and $U_{n\nu}$ is an $SU(N)$ matrix that represents 
the gauge fields. The factors $\eta_{n\nu}$ are the Kogut-Susskind 
factors that give the fermions their Dirac structure
\begin{equation} 
\eta_{n\nu}=\left\{ 
\begin{array}{ll} 
  e^{\mu} & \quad \nu=0, \\ 
  (-1)^{n_0+n_1+\dots+n_{\nu-1}} & \quad \nu\in[1,d], 
\end{array} 
\right.  \label{eq:eta}
\end{equation} 
and here $\mu\ge0$ is the quark chemical potential in
 units of the lattice 
spacing \cite{Hasenfratz_Karsch}. The source
 terms $J,c$, and $\bar{c}$ couple to the following meson and baryon fields 
\begin{eqnarray} 
M_n &=& \frac1N \sum_{a=1}^N \chi_{n,a} \bar{\chi}_{n,a} ,\nonumber \\ 
B_n &=& \chi_{n,1} \chi_{n,2} \cdots \chi_{n,N}, \label{eq:MB}\\ 
\bar{B}_n&=& \bar{\chi}_{n,N} \bar{\chi}_{n,N-1} \cdots \bar{\chi}_{n,1},\nonumber 
\end{eqnarray} 
and as indicated here, the only internal index that $\chi_n$ and 
$\bar{\chi}_n$ carry, is the color index $a=1,\dots,N$. This 
leads to significant simplifications compared to the flavoured case, because it means that $(M_n)^k=0$ for $k>N$, and 
$(B_n)^k=(\bar{B}_n)^k=0$ for $k>1$ \cite{KS,HKS}. 
 
Since the plaquette term is absent
 in \Eq{eq:S_LQCD},
one can readily 
integrate over the $SU(N)$ link matrices. This is performed link by 
link and the result is 
\begin{eqnarray} 
Z(J,c,\bar c)&=&\int D\bar \chi \, D\chi \, DU \exp{S} = \int D\bar \chi \, 
D\chi \, \exp{S_1(M,B,\bar{B})},\nonumber \\ 
S_1&=& \sum_{n\nu}\left\{N F_N(M_{n}M_{n+\hat \nu}) - 
\frac{2^{-N+1}}2\left[ 
  \bar{B}_n \left(\eta_{n\nu}\right)^N B_{n+\hat \nu} - 
  \bar{B}_{n+\hat \nu} \left(\eta_{n\nu}\right)^{-N} B_n \right]\right\}\\ \label{eq:S1}
&&+ \sum_n \left\{N J_n M_n + \bar{c}_n B_n + \bar{B}_n c_n\right\}. \nonumber
\end{eqnarray} 
Here the function $F_N$ is known for
 several values of $N$ \cite{F_N},
 and in particular, for $SU(3)$, it is  
\begin{equation} 
F_3(u)=\frac14 u + \frac{3}{64} u^2 - \frac{15}{256} u^3. \label{eq:F_3}
\end{equation} 
The next step is to write $Z$ as 
a path integral over {\em color-singlet} fields. This is accomplished by
first writing 
\begin{eqnarray} 
Z&=&\exp \left[ S_1(\frac1N \partial_J,
 -\partial_c, \partial_{\bar{c}} ) \right]Z_0,\label{eq:S1_on_Z0} \\
Z_0&=&\int D\bar \chi D \chi
 \exp {\left\{\sum_n\left[NJ_nM_n + 
\bar{B}_n c_n + \bar{c}_n B_n\right]\right\}}.
\end{eqnarray} 
Performing the integral over $\chi$ and $\bar \chi$ results in (here we assume that $N$ is odd)
\begin{eqnarray} 
Z_0&=&\prod_n \left[J^N_n + \bar{c}_n c_n\right],\label{row2}
\end{eqnarray} 
that can be written as 
\begin{equation} 
Z_0=\int D\bar b \,Db \,Dm \,
\exp {\left\{\sum_n\left[ -d_N\bar{b}_n m^{-N}_n 
b_n +NJ_n m_n +\bar{c}_n b_n + \bar{b}_n c_n \right] \right\} },\label{Z0}
\end{equation} 
with $d_N=N!/N^N$. Here $b_n$ and $\bar{b}_n$ are Grassmann variables, and the field
$m_n=e^{i\theta_n}$ takes values on the unit circle and has the measure 
\begin{equation} 
\int dm_n \equiv \oint \frac{dm_n}{2\pi i 
m_n}=\int_{-\pi}^{\pi}\frac{d\theta_n}{2\pi}. \label{eq:m_measure} 
\end{equation}

We are now ready to apply \Eq{eq:S1_on_Z0} and doing so one obtains
 (setting $J=c=\bar{c}=0$, and replacing $b\to-2^{N-1}b$) 
\begin{eqnarray}
Z&=&\int D\bar b \,Db \,Dm \, \exp (S_{\rm HKS}),\\
S_{\rm HKS}&=&S_{\rm Meson}+S_{\rm Baryon}+S_I, \label{eq:Seff}\\
S_{\rm Meson}&=&\sum_{n\nu} N F_N (m_n m_{n+\hat \nu}),\\ 
S_{\rm Baryon}&=&\sum_{nm}\bar{b}_n \,D_{nm} \,b_m, \label{S_B}\\
S_I&=&\sum_{n} \bar{b}_n b_n \, \left( 2^{N-1}d_N m_n^{-N}\right). 
\end{eqnarray}
Here $D$ is the massless Dirac operator of a free one-component
 staggered 
fermion whose chemical potential is equal to $N\mu$, 
\begin{eqnarray} 
D_{n,m} &=& \frac12\left\{ \left[\delta_{m,n+\hat 0}e^{N\mu} 
-\delta_{m,n-\hat 0} e^{-N\mu} \right] + \sum_{\nu=1}^d \eta_{n \nu} 
\left[\delta_{m,n+\hat \nu}-\delta_{m,n-\hat \nu} \right]\right\}.  \label{eq:dirac}
\end{eqnarray} 

Since the composite 
fields $M_n$, $B_n$, and $\bar B_n$ in \Eq{eq:S_LQCD} couple to the same 
currents as $m_n$, $b_n$, and $\bar b_n$ of $S_{\rm HKS}$ (see \Eq{Z0}), 
then their correlation functions are the same. This means
 that the fields $b_n$ and $m_n$ represent the baryons and mesons respectively.
 Their hopping on the lattice is described by $S_{\rm
  Baryon}$ and $S_{\rm Meson}$, while the latter also describes
meson-meson interactions. Interactions between mesons and baryons are
described by the Yukawa-like vertex $S_I$, which 
in terms of quarks,
represents a rearrangement
 of $N$ quark-antiquark pairs into
 a single baryon-antibaryon pair. 

\subsection{Chiral density waves from $S_{\rm HKS}$ ?}
\label{Seff_CDW}

The purpose of this section it is to show why it is natural to expect that chiral density
waves emerge from $S_{\rm HKS}$ (\Eq{eq:Seff}).
We begin with Section~\ref{subsec:CDW_stag} by formulating the chiral wave ansatz of
 Eqs.~(\ref{gs1})-(\ref{gs2}) in terms
of the fields that appear in $S_{\rm HKS}$, and proceed to
Section~\ref{subsec:CDW_mech}, where we explain the mechanism by which
these waves can arise.

\subsubsection{The chiral density wave 
ansatz with staggered fermions}
\label{subsec:CDW_stag}

We begin with constructing the condensates in Eqs.~(\ref{gs1})--(\ref{gs2}). To do so 
we define a new lattice with
 spacing
$a=2$ that has $2^{d+1}$ sites in 
its unit cell (This is the lattice of 
hyper-cubes used to define the taste
 basis \cite{Rothe}). The original lattice
 coordinate $n$ is related to the new one $X$ by
\begin{equation}
n=2X+\rho,
\end{equation}
where the $(d+1)$-dimensional vector $\rho$ has
 $\rho_\nu=0,1$, and denotes 
the internal sites in the unit cell.
 The mean-field ansatz that 
we study in this paper is
\begin{equation}
  \<\left( \frac{\bar \chi_n \chi_n}{N} \right)^q\>=
\<\left(m_n \right)^q\>=V_q \, \,
 e^{iq\,  \vec Q\,  \vec n \,\epsilon_n} , \qquad q=1,2,\dots,N.\label{ansatz0}
\end{equation}
where the sign factor $\epsilon_n$ is 
given by
\begin{equation}
\epsilon_n=\left(-1\right)^{\sum_{\nu=0}^{d} n_\nu}=\left\{ 
\begin{array}{lr}
+1      & \,{\rm even \,\,site}\\
-1              & \,{\rm odd \,\,site}
\end{array}
\right..
\end{equation}
The appearance of this sign factor 
in the phases of the condensates realises the 
helical structure of Eqs.~(\ref{gs1})--(\ref{gs2}) in the staggered 
formalism. To see this, we transform to taste basis
 with the  unitary
transformation $U$ \cite{Rothe} 
\begin{eqnarray}
\chi_\rho(X)&\equiv&\sum_{\alpha=1}^{2^{D/2}}\sum_{f=1}^{2^{D/2}}
\left( U^\dag\right)_{\rho,(\alpha,f)} \psi_{\alpha,f}(X),\label{taste1}\\
\bar \chi_\rho(X)&\equiv&\sum_{\alpha=1}^{2^{D/2}}\sum_{f=1}^{2^{D/2}}
\bar \psi_{\alpha,f}(X) U_{(\alpha,f),\rho},\label{taste2}\\
U_{(\alpha,f),\rho}&=&{\cal N}_0
\left( \prod_{\nu=1}^{D} \gamma_\nu^{\rho_\nu} \right)_{\alpha,f}.\label{taste3}
\end{eqnarray}
Here $D=d+1$ is the number of spacetime dimensions, which we restrict
to $D=1+1$ or $D=3+1$. The indices $\alpha$ and $f$ are
 identified with the Dirac and taste indices respectively, and 
 $(\alpha,f)$ is a composite index that takes $2^D$ values. The normalisation 
${\cal N}_0$ is a chosen to have
 $U^\dag U = \bm{1}$, and the matrices
 $\gamma_{\mu}$ are the Euclidean Dirac 
matrices.\footnote{In $d=1$ we choose $\left(\gamma_0,\gamma_1,\gamma_5\right)$ to be the Pauli matrices $\left(\sigma_z,\sigma_y,\sigma_x\right)$, and in $d=3$ we use the convention of \cite{Rothe}.}  Using Eqs.~(\ref{taste1})--(\ref{taste3}) and the fact that the sign factor 
$\epsilon_n$ depends only on
$\rho$,
\begin{equation}
\epsilon_n=\epsilon_{2X+\rho}=\left(-1\right)^{\sum_{\nu}\rho_\nu} \equiv\epsilon_\rho,\label{epsrho}
\end{equation}
 we can write 
\begin{eqnarray}
\sum_\rho\<\bar \chi_\rho(X) \chi_\rho(X)\> &=&\sum_{\alpha,f}\<\bar \psi_{\alpha,f}(X) \psi_{\alpha,f}(X)\>\equiv \<
\bar \psi(X) \psi(X)\>,\label{helical1}\\
\sum_\rho\<\bar \chi_\rho(X) \,i \epsilon_\rho\, \chi_\rho(X) \>&=&i\sum_{\alpha,f\atop \alpha',f'}\< \bar \psi_{\alpha,f}(X) \left(\gamma_5\right)_{\alpha\alpha'} \left(\xi_5\right)_{ff'} \psi_{\alpha',f'}(X)\>\nonumber \\ 
&\equiv& \<\bar \psi(X) \,i\,\left( \gamma_5 \otimes \xi_5\right) \psi(X)\>.\label{helical2}
\end{eqnarray}
Here $\gamma_5$ and $\xi_5$  have the same matrix elements, and act in
Dirac and taste space respectively.
Substituting \Eq{ansatz0} in the left hand side of
Eqs.~(\ref{helical1})--(\ref{helical2})
and using \Eq{epsrho} gives
\begin{eqnarray}
\<\bar \psi(X)\, \psi(X)\>&=&  A \,\cos \left( 2\vec Q \cdot \vec
  X + \phi \right),\label{gs1X} \\
\<\bar \psi(X) \left( i \,\gamma_5 \otimes \xi_5 \right) \psi(X)\>&=&-
A \,\sin \left( 2\vec Q \cdot \vec X + \phi \right),\label{gs2X}
\end{eqnarray}
with 
\begin{eqnarray}
\phi&\equiv& \frac12 \sum_{\nu=1}^d Q_\nu,\\
A&\equiv&2 V_1 \left\{
\begin{array}{lr}
\cos \left(\frac12 Q_1\right) & \quad d=1 \\
\cos \phi+ \sum_{\nu=1}^3 \cos \left( \frac12
    Q_\nu -\frac12 \sum_{\mu\neq \nu} Q_\nu\right) & \quad d=3
\end{array}
\right.,
\end{eqnarray}
By performing the chiral rotation of \Eq{U1eps} with
$\theta=\phi/2$ we get rid of 
the angle $\phi$ in the arguments of the
sine and cosine functions in Eqs.~(\ref{gs1X})--(\ref{gs2X}), and 
obtain the assured helical structure of
Eqs.~(\ref{gs1})--(\ref{gs2}).

\subsubsection{The mechanism of the chiral
 density wave instability}
\label{subsec:CDW_mech}

We now proceed to show that the Yukawa-like interaction, $S_I$,
 can 
make the state characterised by 
\Eq{ansatz0} become the 
ground state of $S_{\rm HKS}$ (\Eq{eq:Seff}) at $\mu>0$. For 
 simplicity, we consider the $d=1$ case only. 

In the absence of $S_I$, the baryons behave like free massless
fermions, and it is straight-forward to 
show that the poles of the propagator 
$D^{-1}$ of \Eq{S_B} are determined in momentum space by
\begin{equation}
0=\left[\sin^2 \left( p_0/2 - i \mu N \right)
+  \sin^2 \left(p_1/2 \right)\right]^2\quad ; \quad -\pi\le p_{0,1} \le +\pi. 
\label{zeroth1}
\end{equation}
 \Eq{zeroth1} tells us that $S_{\rm Baryon}$ describes the four
 energy bands
\begin{eqnarray}
\sinh E^{(1),(2)}_0(p_1)&=&+|\sin(p_1/2)|,\label{zeroth}
\\ 
\sinh E^{(3),(4)}_0(p_1)&=&-|\sin(p_1/2)|, \label{zerotha}
\end{eqnarray}  
with $p_1 \in [-\pi,+\pi]$, and that
 the Fermi energy and Fermi momentum of
the baryons are 
\begin{eqnarray}
E_F&=&\mu N, \label{EFermi}\\
\sin p_F/2&=& \sinh N \mu.\label{Fermi}
\end{eqnarray}

To take $S_I(b,\bar b, m)$ into account we use
 mean-field
theory, and replace the meson fields by their condensates.  
In the next section we find that with the mean-field ansatz of
\Eq{ansatz0} we should replace $S_I$ with
\begin{equation}
S^{\rm mean-field}_I \equiv \sum_n \bar b_n b_n \left( \Sigma \, e^{-i N Q\, n_1\, \epsilon_n}\,\right) ,\label{mn}
\end{equation}
where the amplitude $\Sigma$ is a complicated function of
the amplitudes $V_{1,2,\dots,N}$ from \Eq{ansatz0}.
 In terms of the lattice coordinates of the new lattice \Eq{mn} becomes
\begin{equation}
S^{\rm mean-field}_I=\sum_{X\rho} \bar b_\rho(X) b_\rho(X) \left(
  \Sigma e^{-i N\,  Q\, (2 X_1+ \rho_1)  \, \epsilon_\rho}\right), \label{mX}
\end{equation}
where we defined $b_n=b_{2X+\rho}\equiv b_\rho(X)$ and similarly for $\bar b$.

\begin{figure}[htb]
\includegraphics[width=10cm]{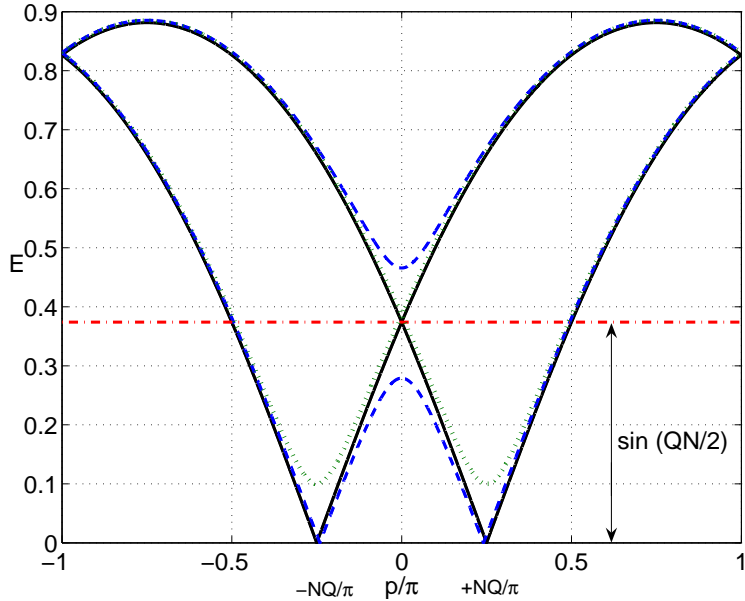}
\caption{Dispersion relations  $E_\pm$ : 
  $\Sigma=Q=0$ (black solid lines), 
$\Sigma=0.1$ and $NQ=\pi/4$ (dashed blue lines), 
and $\Sigma=0.1$, and $Q=0$ (dotted green lines). 
Note that the solid and dotted lines have been shifted $\pm QN$
from the origin, so that they have the same abscissa as the
 dashed lines. With these shifts the level repulsion occurs at
$p_1=0$, where the $\Sigma=0$ energy bands are degenerate with $E_+=E_-=\sin
(QN/2)$. 
\label{fig:sketch}}
\end{figure}

Clearly, $S^{\rm mean-field}_I$ mixes baryons 
whose momenta differ by the amount 
$\delta p_1=2NQ$. The strongest
 effect will occur
between baryons that in the absence of $S_I$ are degenerate in energy, and in particular, this
mixing can occur between the bands $E^{(1)}$ and
$E^{(2)}$ of \Eq{zeroth}.
 This will lead to level repulsion,
 and to the opening of a gap in the 
spectrum. To see this explicitly we move to
momentum space with\footnote{Although we insert 
identical phases of $\epsilon_\rho NQ (2X_1+\rho_1)/2$ into \Eq{FT1} 
and \Eq{FT2}, the 
transform between
 $X$-space and $p$-space still has a 
unit determinant due to the sign factor 
$\epsilon_\rho$. (On a lattice with an even number of sites in each direction).}
\begin{eqnarray} 
b_\rho(X)&=&\sqrt{\frac{4}{N_s}} \, 
e^{-i\epsilon_\rho 
N Q(2X_1+\rho_1)/2} \,\sum_{p}
 e^{ + i p X} \,b_\rho(p),\label{FT1} \\
\bar b_\rho(X)&=&\sqrt{\frac{4}{N_s}}
\, 
e^{-i\epsilon_\rho 
N Q(2X_1+\rho_1)/2} \, \sum_{p} e^{-i p X} \,\bar b_\rho(p). \label{FT2}
\end{eqnarray} 
Note that Eqs.~(\ref{FT1})--(\ref{FT2}) are
 not a usual Fourier transform,
 because they gives the field $b_\rho(p)$ a momentum that differs 
from the momentum of $\bar b_\rho(p)$ by an amount $\delta p_1 = 2NQ \epsilon_\rho$. 
This makes $S_I^{\rm mean-field}$
 diagonal in $p$ space, 
and we find that the contributions of the baryons to $S_{\rm HKS}$ are given by
\begin{eqnarray}
S_{\rm Baryon} + S^{\rm mean-field}_I&\equiv&\sum_{p\atop \rho\rho'}\bar{b}_\rho(p)\, K_{\rho\rho'}(p) \,b_{\rho'}(p),
\end{eqnarray}
where the Dirac operator $K$ is
\begin{eqnarray}
K_{\rho\rho'}(p)&=& \left[ \Gamma_0 \, \sin \left(p_0/2-i\mu N\right) + \Gamma_1 
\sin\left(p_1/2 + \hat \epsilon\,
  QN/2\right)    + \Sigma
\, \bm{1} \right]_{\rho\rho'}. \label{SI_mix}
\end{eqnarray}
Here 
\begin{eqnarray}
\left(\hat \epsilon\right)_{\rho\rho'}&\equiv& \epsilon_\rho\delta_{\rho\rho'},\\
\left(\Gamma_\nu\right)_{\rho\rho'}&\equiv&
\tilde \eta_\nu\left( \delta_{\rho',\rho+\hat \nu} + \delta_{\rho',\rho-\hat \nu}
\right)e^{ip\left(\rho-\rho'\right)/2},
\end{eqnarray}
with $\tilde \eta_0=1$ and $\tilde \eta_1=(-1)^{\rho_0}$.
Using the methods of
Appendix~\ref{sec:baryon_det}, it is
 easy to calculate the
determinant of $K(p)$, extract its roots, 
and find that the four energy levels in Eqs.~(\ref{zeroth})--(\ref{zerotha}) are modified
 to two positive energies,
\begin{eqnarray}
\sinh E_\pm &=& +\left|\sqrt{\Sigma^2+\left(\sin (p_1/2) \cos (NQ/2)\right)^2} \pm \cos (p_1/2) \sin (NQ/2) \right|,\label{level_rep}
\end{eqnarray}
and two negative energies 
$\bar E_\pm = - E_\pm$. 
 We plot $E_\pm$ in Fig.~\ref{fig:sketch}, 
 where we choose $\Sigma=Q=0$ (black solid lines), 
$\Sigma=0.1$ and $NQ=\pi/4$ (dashed blue lines), and $\Sigma=0.1$ and
$Q=0$ (dotted green lines). Note that the case with $\Sigma=0$ corresponds to the 
absence of $S_I$ and the band structure should not depend
on $Q$. Indeed, from \Eq{level_rep}
we see that if $\Sigma=0$ the energies $E_\pm$ reduce to $E^{(1),(2)}$
(up to a shift of $\pm NQ$ from the origin, that can be removed by
redefining the Brillouin Zone).

Next, consider the following choice of the wave-vector $Q$
\begin{equation}
NQ = p_F.\label{Q_pF}
\end{equation}
Together with \Eq{Fermi} and \Eq{zeroth}, this  
identifies the dashed-dotted horizontal line of
Fig.~\ref{fig:sketch} with the Fermi level \Eq{EFermi}, 
in which case the effect of
the level repulsion is to lower all the energies in the 
Fermi sea (see dashed blue
line in Fig.~\ref{fig:sketch}).
 This {\em decreases} the baryon contribution to the free energy of the
system with respect to the $\Sigma=0$ case.
 In contrast, if we choose $Q=0$ then
 $S_I$ is a simple mass
term that makes
\begin{equation}
\sinh E_\pm =\sqrt{\Sigma^2 + \sin^2(p_1/2)},\label{massive}
\end{equation}
 and
{\em increases} the energy of the
 Fermi sea (see dotted green
line in Fig.~\ref{fig:sketch}). As a result, 
 if we
denote the contribution of the baryons to the free energy by ${\cal  E}_B(\Sigma,Q)$ then
\begin{equation}
{\cal E}_B(\Sigma,p_F/N) < {\cal E}_B(0,0) < {\cal E}_B(\Sigma,0), \label{delE}
\end{equation}
and so, for $\mu>0$, the Fermi sea energetically prefers that the
 condensates in Eqs.~(\ref{ansatz0}) carry a nonzero
momentum Q.

In the discussion above we have ignored many
 details. Firstly, we addressed 
only $d=1$, where
 both the Fermi sphere
and the gap occur at single points that
can always be made to coincide. For $d>1$,
 the Fermi sphere is a curved
surface, and it is not assured that one can
 find a wave $\vec Q$ that
will change the energy bands such
 that all the energies in the Fermi sea
are pushed down.
 Indeed, it is known in condensed matter physics that similar
 inhomogeneous 
 instabilities take place for $d>1$ only for sufficiently strong interactions
 \cite{Peierls,Overhauser}. 
Secondly, we discussed only the contribution of the Fermi sea to the
free energy. The remaining contributions are the anti-baryons energy
(coming in the form of the negative energy bands $\bar E_\pm$ of the
Dirac sea) and the meson self-energy coming from $S_{\rm
   Meson}$. In fact, as we find in the next sections, both 
these contributions depend on $Q$, and 
can prefer $Q=0$ for all values of $\mu\ge 0$.
 Thirdly, $\Sigma$ may 
also depend on $Q$, while in the
discussion above we have treated it as being fixed. 

Finally, note that our discussion relies on the level repulsion
presented in Fig.~\ref{fig:sketch}, where we choose $\Sigma=0.1$. 
Since $S_I$
also gives mass to the baryons (see \Eq{massive}), this means
that we have, so far, had in mind
small baryon masses. It is a priori unclear 
whether the chiral wave instability occurs 
for large values of $\Sigma$ and very massive baryons, 
but what is clear
is that if 
the energy bands of the baryons in the crystalline phase 
look qualitatively like what we present in
Fig.~\ref{fig:sketch}, then it is 
 natural to expect this instability
 when $NQ\simeq p_F$. More precisely, by comparing 
the solid black lines and the dashed blue lines in 
Fig.~\ref{fig:sketch}, we see that 
what allows the crystalline phase (that has nonzero 
$\Sigma$ and $Q$) to compete with the massless phase 
(that has $\Sigma=0$) is the fact that it has gapless excitations.

To study the effects of the issues we mention above on the 
viability of the chiral density waves, we develop
a mean-field analysis for inhomogeneous vacua
 in the next Section.

\section{Mean-field theory} 
\label{sec:MFT} 
 
We start from \Eq{eq:Seff} and introduce auxiliary fields for 
the expectation values of $m_n, (m_n)^2,\dots,(m_n)^N$ by 
writing \cite{ZJ}
\begin{eqnarray} 
\exp \left[ N\sum_{n\nu}F_N(m_n m_{n+\hat \nu}) \right]&=&\exp \left[N\sum_{q=1}^N\sum_{n\nu} 
  a_q m^q_n m^q_{n+\hat \nu}\right] \nonumber \\ 
&=& \int DV \exp \left[N\sum_{q=1}^N \sum_{n\nu} a_q V_{q,n} 
  V_{q,n+\hat \nu}\right] \delta (V_{q,n}-m^q_{n}) \label{eq:VH_tranf} \\
&=& \int DV Dh \exp \left[N\sum_{q=1}^N  \sum_{n\nu} a_q V_{q,n} 
  V_{q,n+\hat \nu} - N h_{q,n} \left(V_{q,n}-m^q_n\right)\right].\nonumber 
\end{eqnarray} 
Here $\displaystyle{\int dV \equiv 
  \prod_q \int_{-\infty}^\infty dV^{(q)}_{R} \int_{-\infty}^\infty dV^{(q)}_{I}}$, and $V^{(q)}_{R,I}$ 
are the real and imaginary parts of $V_q$, while $\displaystyle{\int 
dh \equiv \prod_q \int_{-i\infty}^{i\infty} dh^{(1)}_q \int_{-i\infty}^{i\infty}
dh^{(2)}_q}$. Also, $h_qV_q$ means $h^{(1)}_{q} V^{(q)}_{R} + h^{(2)}_{q} V^{(q)}_{I}$, 
and similarly for $h_q(m)^q$.  
Proceeding from \Eq{eq:VH_tranf}, the action for the scalar fields $V$
and $h$ becomes
\begin{eqnarray} 
S_{\text eff}(V,h) &=&N\sum_{q=1}^N a_q \sum_{n\nu} 
V_{q,n}V_{q,n+\hat \nu} - N \sum_{n,q} h_{q,n} V_{q,n}  + S_0, \label{eq:SeffVH} \\ 
\exp (S_0)&=&\int D\bar b \, Db \, Dm \, \exp{ \left\{ \, \bar{b} \cdot\left[ 2^{N-1}
    d_N m^{-N} \bm{1} + D \right]\cdot\,b + N \sum_{q,n} h_{q,n} m^q_n \right\}}.\label{eq:expS0} 
\end{eqnarray} 
 
The starting point of Mean field theory is to
 write the path integral for the action in
\Eq{eq:SeffVH} as
\begin{equation} 
Z(\lambda)\equiv\int Dh \, DV \, e^{\frac1{\lambda} S_{\text 
eff}} \label{Z_lam}
\end{equation} 
 and to approach the physical point, $\lambda=1$, by expanding around
 $\lambda=0$  \cite{ZJ}. Calculating $O(\lambda)$ corrections
 is, however, outside the scope of the work we present here, and we study only the
 $O(\lambda^0)$ level. In this case, the tree-level mean-field equations are
\begin{equation} 
\frac{\partial S_{\text eff}}{\partial V_{q,n}}=0, \qquad {\text and} \qquad \frac{\partial 
S_{\text eff}}{\partial h^{(1,2)}_{q,n}}=0,
\end{equation} 
which give
\begin{eqnarray} 
h^{(1)}_{q,n} &=& a_q \sum_{\nu=\pm 0}^{\pm d} V_{q,n+\hat \nu}, \label{eq:MFh1}\\ 
h^{(2)}_{q,n} &=& i a_q \sum_{\nu=\pm 0}^{\pm d} V_{q,n+\hat \nu}, \label{eq:MFh2}\\ 
V_{q,n} &=& \<m_n^q\>. \label{eq:MFV}
\end{eqnarray} 
Here the average $\<,\>$ in \Eq{eq:MFV} is with respect to the path integral in \Eq{eq:expS0}.

 Equations~(\ref{eq:MFh1})-(\ref{eq:MFh2}) tell us to try an
ansatz with $h^{(2)}=ih^{(1)}\equiv
ih$, in
which case the $h V$ and $h m$ terms in 
\Eq{eq:SeffVH} and \Eq{eq:expS0} have the  meaning of a simple complex
multiplication. Restricting to this type of ansatze we use
\begin{equation} 
\oint \frac{dm}{2\pi i m} m^k = \delta_{k,0}, 
\end{equation} 
to perform the ${\displaystyle \int Dm}$ integral in \Eq{eq:expS0}, and we find (for $N=3$)
\begin{equation} 
\oint \frac{dm}{2\pi i m} \exp \left[ 3\left(h_1 m +  h_2  m^2 + h_3 m^3 \right)
+\frac89 m^{-3} \bar{b}b \right] = \exp 
\left[4\left(h^3_1+2h_1h_2+\frac23 h_3\right)\bar{b}b\right].
\end{equation} 
This turns $Z(1)$ of \Eq{Z_lam} to 
\begin{eqnarray} 
Z&=&\int DV  \, Dh \, D\bar b \, Db \, \exp S_{\text eff},\\ 
S_{\text eff} &=&N\sum_{q=1}^N a_q \sum_{n\nu} 
V_{q,n}V_{q,n+\hat \nu} - N \sum_{n,q} h_{q,n} V_{q,n} + \sum_{n,m}\bar{b}_n \left[ 
  \Sigma_n \, \bm{1} +  D \right]_{nm}b_m,\\ \label{eq:SMF} 
\Sigma_n&=&4\left(h^3_{1,n}+2h_{1,n}h_{2,n}+\frac23 h_{3,n}\right).
\end{eqnarray} 
 
The next step is to {\em assume} an ansatz for the spatial 
behaviour of $V_{q,n}$ and $h_{q,n}$. Here we allow for the 
possibility of the broken translation invariance, and write 
\begin{equation} 
V_{q,n}  =  V_q \,\exp\left( +iq\, \epsilon_n 
\,\vec Q \,\vec n\right), \qquad
h_{q,n}  =  h_q \,\exp\left( -iq\, \epsilon_n 
\,\vec Q \,\vec n\right), \label{eq:ansatz}
\end{equation}
with real $h_q$ and $V_q$. 
This ansatz breaks the symmetry of spatial translations as well as the $U(1)_\epsilon$ chiral rotations 
\begin{equation} 
V_{q,n}\to V'_{q,n}=e^{i\,\epsilon_n \,q\,\theta}\, V_{q,n}, \qquad 
h_{q,n}\to h'_{q,n}=e^{-i\,\epsilon_n \,q\,\theta}\, h_{q,n}, \label{eps_sym}
\end{equation} 
and, as we discuss in Section~\ref{Seff_CDW}, implies a helical
structure for the chiral condensate.

A substitution of \Eq{eq:ansatz} in 
\Eq{eq:MFh1} gives 
\begin{eqnarray}
h_q&=&2(d+1)a_q V_q \gamma_{q}(Q),\label{eq:MFhV}
\end{eqnarray}
where 
\begin{eqnarray}
 \gamma_q(Q)&\equiv&\frac{1}{d+1}\left[1+\sum_{i=1}^d \cos \left(q Q_i\right)\right], \label{gammaQ}
\end{eqnarray}
and a subsequent 
substitution of that into \Eq{eq:MFV}, with a use of 
\begin{equation} 
\frac{\partial \Sigma_n}{\partial h_{nq}}=4\times \left\{\begin{array}{lr} 
3h^2_{1,n}+2h_{2,n} & \quad q=1 \\ 
2h_{1,n} & \quad q=2 \\ 
\frac23  & \quad q=3 
\end{array} 
\right. ,
\end{equation} 
gives the final form of the mean-field 
equations
\begin{eqnarray}
h_1&=&\frac92 \frac{a_1\gamma_1}{a_3\gamma_3}\left(h^2_1 +  \frac{2a_2 
    \gamma_2}{a_3 \gamma_3}h_1h_3 \right)h_3,\label{eq:MFeq1} \\ 
h_2&=& \frac{3a_2\gamma_2}{a_3\gamma_3} \, h_1h_3,\label{eq:MFeq2}  \\
h_3&=&\frac{16(d+1)a_3\gamma_3}{9}\times A(h_q,Q;\mu). \label{eq:MFeq3}
\end{eqnarray} 
Also, $a_{1,2,3}=\frac14,\frac{3}{64},-\frac{15}{256}$, are the coefficients 
in $F_3$ (see \Eq{eq:F_3}), and $\gamma_{1,2,3}$ are given in \Eq{gammaQ} (Here, for brevity, we write $\gamma_q$ to denote $\gamma_q(Q)$).
The function $A(h_q,Q;\mu)$ is defined as
\begin{equation} 
A(h_q,Q;\mu)\equiv \<\bar{b}_nb_n\>\cdot e^{-3i\vec Q \vec n\epsilon_n} =\frac{\partial \, \tr \log \left[\Sigma_n\bm{1} + D 
\right]}{\partial\left( \Sigma_n e^{3i\vec Q\vec n}\right)}, \label{eq:A}
\end{equation} 
and the derivative here is evaluated at 
\begin{eqnarray}
\Sigma_n&=&\Sigma  e^{-3i\vec Q\vec n},\\
\Sigma&=&4\left(h^3_{1}+2h_{1}h_{2}+\frac23 h_{3}\right). \label{eq:Sigma}
\end{eqnarray}
 To determine which of the solutions of the mean-field equations is the 
true vacuum we calculate the mean-field free 
energy. Using \Eq{eq:MFhV}, the energy 
per number of sites $N_s$ is given by
can be expressed in terms of the fields $h_q$ alone, and
we find
\begin{eqnarray} 
{\cal E}&\equiv& -S^{\rm mean-field}_{\rm eff}/N_s=\frac34 \sum_q \frac{h_q^2}{(d+1)a_q\gamma_{q}(Q)}+ {\cal 
E}_{\text{matter}},\label{eq:EMF}
\end{eqnarray} 
where the contribution of the $b$ fields is 
\begin{eqnarray} 
{\cal E}_{\text{matter}}&=& -\frac{1}{N_s} \tr \log \left[\Sigma_n 
\bm{1} + D \right]. \label{eq:det}
\end{eqnarray}

We calculate $A(h_q,Q;\mu)$ and ${\cal E}_{\text matter}$ in Appendix~\ref{sec:baryon_det}, where we also show that
the right hand side of \Eq{eq:A} is indeed independent of the lattice
site index $n$. The result for ${\cal
  E}_{\text matter}$ is
\begin{equation} 
{\cal E}_{\rm matter} = -\frac12 \sum_{b=\pm} \int 
\left(\frac{dp}{2\pi}\right)^{d} \left( 
  E_b-3\mu \right) \theta \left( 
  E_b -3\mu\right), \label{Ematter}
\end{equation} 
where $\theta(x)$ is the step function.
 The momentum integrals are 
from $-\pi$ to $\pi$, and $E_\pm$,
 that replace $E^{(1),(2)}$ of 
\Eq{zeroth}, are 
the band energies of the baryons.
In this paper we
focus on $d=3$ and $\vec Q=Q\hat z$, for which 
the dispersions are
\begin{eqnarray} 
\sinh^2 E_\pm &=& \sin^2 (p_x/2)+\sin^2 (p_y/2) + \left(
  \sqrt{\Sigma^2+s^2_z} \pm 
|c_z| \right)^2, \label{eq:Epm} \\
s_z&\equiv&\sin (p_z/2)\cos 3Q/2,\\
c_z&\equiv& \cos (p_z/2)\sin 3Q/2. 
\end{eqnarray} 
(For the general dispersions with $\vec Q=(Q_x,Q_y,Q_z)$ see
Appendix~\ref{sec:baryon_det}).
$\Sigma$ is related to the expectation values $V_q=\<m^q_n\>$ through
Eqs~(\ref{eq:MFh1})--(\ref{eq:MFV}) and for $Q=0$ it is the spectral
gap that defines the baryon mass $m_B$ (see \Eq{eq:Epm})
\begin{equation} 
\sinh m_B = \Sigma.\label{mBdef}
\end{equation}
From here on we refer to
energy bands of \Eq{eq:Epm} 
with $\Sigma>0$ as massive bands, 
and to energy bands with $\Sigma=0$ 
as massless bands.
 
\begin{figure}[htb]
\includegraphics[width=13cm]{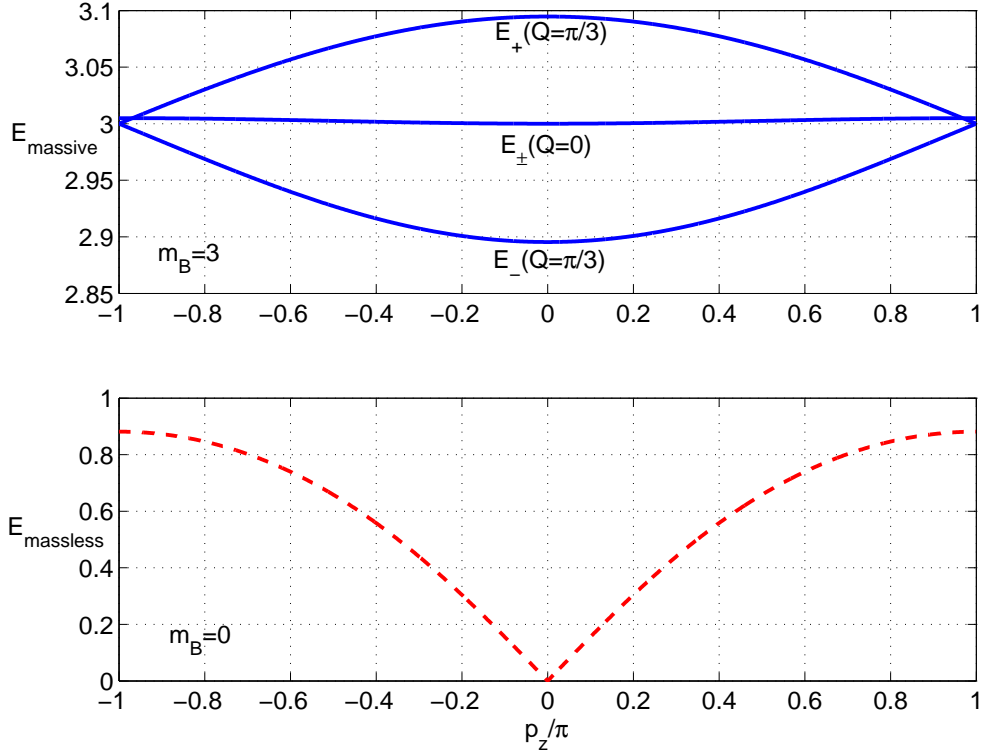}
\caption{The energy bands $E_\pm$ of \Eq{eq:Epm}
for 
  $p=(0,0,p_z)$. \underline{Upper panel:} Massive bands with $m_B=3$ and
  $Q=\pi/3$ (highest and lowest curves), and $Q=0$ (middle curve).
  The bands $E_\pm(Q=0)$ are degenerate and very flat, ranging between
  $E=3$ and $E=\sqrt{\sinh^2(3)+{\text max}\left( \sum_{i=1}^3
      \sin^2(p_i/2) \right)}\simeq 3.0147$. \underline{Lower panel:} The
  massless band with
 $\Sigma=0$. 
\label{fig:Epm_N3}}
\end{figure}

In Fig.~\ref{fig:Epm_N3} we plot 
$E_\pm$  for
 $\vec p=(0,0,p_z)$. 
In the upper panel we choose $m_B=3$
 (or $\Sigma\sim 10$), which is close 
to the value that solves the mean-field 
equations in $d=3$ and $\mu=0$ (see next section). We 
present the cases of $Q=0$, where $E_\pm$ are degenerate, 
and $Q=\pi/3$, where this degeneracy is removed. In the lower panel 
we show
the case of $m_B=0$ (or $\Sigma=0$), where $E_\pm$ are again degenerate 
and also independent of $Q$. 

Finally, to make a connection with the discussion 
in Section~\ref{Seff_CDW} leading 
to \Eq{delE}, 
we write \Eq{Ematter} as 
\begin{equation} 
{\cal E}_{\text matter}=-|{\cal E}_{\bar{B}}| + {\cal E}_B - 3 \mu \,
n_B, 
\end{equation} 
where the anti-baryon and baryon energies, $-|{\cal E}_{\bar{B}}|$ and
${\cal E}_{B}$, are given by
\begin{eqnarray} 
-|{\cal E}_{\bar{B}}|&=&+3\mu - \frac12
\sum_{b=\pm} \int 
\left(\frac{dp}{2\pi}\right)^{d} E_b,\\ 
{\cal E}_B &=& \frac12 \sum_{b=\pm} \int 
\left(\frac{dp}{2\pi}\right)^{d} \left( 
  E_b-3\mu \right) \theta \left( 
  3\mu-E_b \right),
\end{eqnarray} 
and the baryon number density $n_B$ is
\begin{eqnarray} 
  n_B&=&\frac12 \sum_{b=\pm} \int 
\left(\frac{dp}{2\pi}\right)^{d} \theta \left( 
  3\mu-E_b \right).
\end{eqnarray} 

\section{The Mean-field ground state} 
\label{sec:MFsol} 
 
In this section we present the solutions for the mean-field equations 
Eqs.~(\ref{eq:MFeq1})--(\ref{eq:MFeq3}), and their corresponding energy densities. To
understand the results we then study a formal limit of
\Eq{eq:Seff} where one takes $N\to 1$.

\subsection{Physical case : $N=3$ and $d=3$} 
\label{sec:physical}

 We begin by describing the solution to the mean-field equation 
that leaves chiral symmetry intact in
 Section~\ref{subsec:SolI}, and proceed to discuss the chiral broken phase in 
Section~\ref{subsec:SolII}. Finally, 
in Section~\ref{subsec:evolve} we discuss how the ground state evolves with increasing values of $\mu$.
\subsubsection{Solution I : Intact chiral symmetry} 
\label{subsec:SolI} 

\begin{figure}[htb]
\includegraphics[width=10cm]{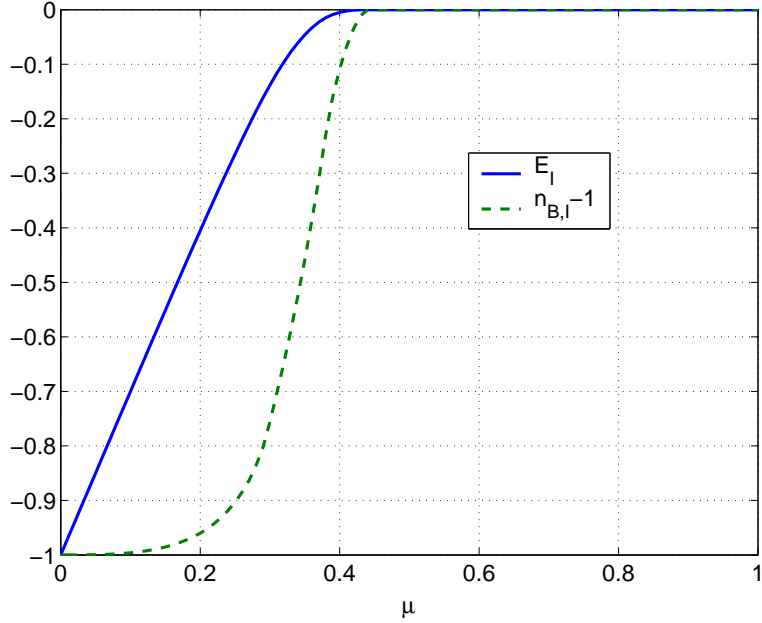}
\caption{The energy density, ${\cal E}_I$, (solid line) and baryon number
  density, $n_{B,I}$, for the massless energy bands of solution type I, as a function of the chemical potential $\mu$.
\label{fig:ENI}}
\end{figure}

The first solution to the mean-field equations is obtained by taking $h_1=h_2=h_3=0$. This 
 corresponds to a phase with intact chiral symmetry, $V_1=V_2=V_3=0$, and 
zero baryon mass $\Sigma=\sinh( m_B)=0$. In this
 case the energy 
bands are trivially 
independent of $Q$ (since $S_I^{\rm mean-field}=0$), and 
$E_\pm$ are degenerate
\begin{equation} 
E_\pm =\sinh^{-1} \sqrt{ \sum_{\nu=1}^d \sin^2 (p_\nu/2) }\equiv E_{\rm massless}. \label{Egapless}
\end{equation} 
The energy and baryon number densities of this solution are independent of $Q$ as well, and
are given by
\begin{eqnarray} 
{ \cal E}_I &=& -  \int \left( \frac{dp}{2\pi}\right)^3 \left( 
  E_{\rm massless}-3\mu \right) \theta \left( 
  E_{\rm massless}-3\mu \right), \\
n_{B,I} &=& \int \left( \frac{dp}{2\pi}\right)^3 \theta \left(3\mu -E_{\rm massless}\right). 
\end{eqnarray} 
In Fig.~\ref{fig:ENI} we plot ${\cal E}_I$ as a function of $\mu$, where one can see 
that for $\mu=0$, we get ${\cal E}_I\simeq -1$. As we increase $\mu$,
the energy density increases, and saturates when $\mu=\mu_s\simeq
0.439$, where it reaches zero. 
Looking at the form of $E_{\rm
  massless}$ from Fig.~\ref{fig:Epm_N3}, we can understand why. The maximum energy to which
$E_{\rm massless}$ of \Eq{Egapless} can reach is $\sinh^{-1} \sqrt{\max
  \sum_{i=1}^3 \sin^2(p_i/2)}\simeq 1.317$. This means that when
$\mu>\mu_s$, where the baryon chemical potential $3\mu$ is already larger than
$3\times 0.439\simeq 1.317$, the band $E_{\rm massless}$ is already saturated.
This saturation can be seen also in the
plot of the baryon number density, $n_{B,I}$
which reaches 
$n_{B,I}=1$ at $\mu_s$. 

\subsubsection{Solution II : Spontaneously broken chiral symmetry} 
\label{subsec:SolII} 

In this case $h_{1,2,3}$ are nonzero, and we divide \Eq{eq:MFeq1} 
with $h_1$ to obtain
\begin{equation} 
h_1=\frac29 \frac{a_3 \gamma_3}{a_1 \gamma_1}\frac1{h_3} -  \frac{2a_2 
  \gamma_2}{a_3 \gamma_3}h_3. 
\end{equation} 
Substituting this in \Eq{eq:MFeq2} 
and \Eq{eq:Sigma} we get $\Sigma$ as a
 function of $h_3$, 
which we then substitute into 
\Eq{eq:MFeq3} and obtain a single 
equation for $h_3$, that we solve for all
values of $Q$ and $\mu$. 

\begin{figure}[htb]
\includegraphics[width=10cm]{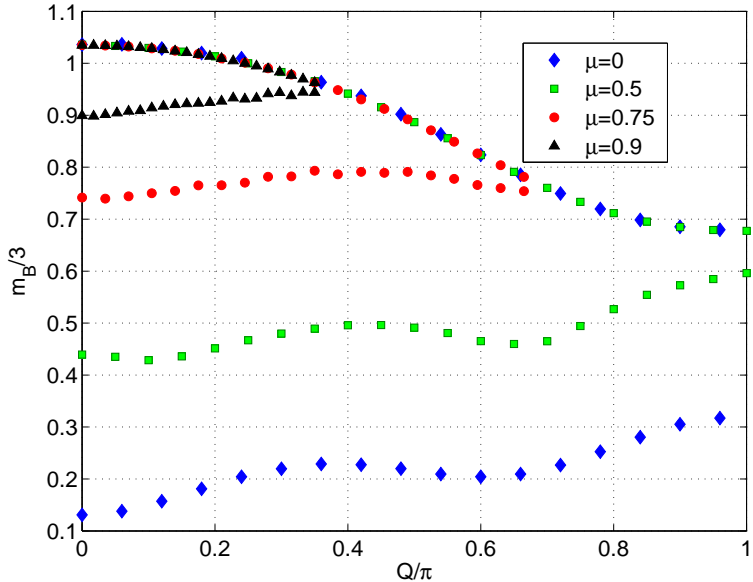}
\caption{The values of
$\frac13 m_B(Q;\mu)$ that solve the mean-field
  equations. Note the two branches of the solution, and that one of
  them (branch no.1, ending at $Q=0$, and $m_B/3=1.0347$) hardly changes with
  $\mu$. We find that this branch has a lower energy and therefore
  corresponds to the ground state (see text).
\label{fig:resultsII}}
\end{figure}

\begin{figure}[htb]
\includegraphics[width=10cm]{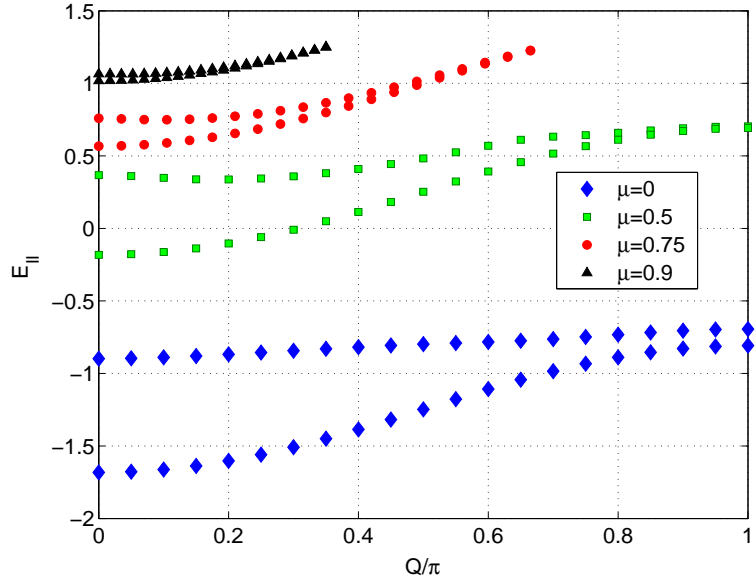}
\caption{The energy density 
${\cal E}_{II}$  as a function of $Q$ and $\mu$ for solution II. The
lower branch (branch no.1) for each $\mu$ is the one that ends at $Q=0$ and
$m_B/3=1.0347$ (see Fig.~\ref{fig:resultsII}). 
\label{fig:ENII}}
\end{figure}

 The results are presented in
 Fig.~\ref{fig:resultsII} where we show
 $m_B\equiv \sinh^{-1} (\Sigma)$ as a function 
of $Q$, for $\mu=0,0.5,0.75,0.9$. 
In Fig.~\ref{fig:ENII} we show the 
energy densities of 
these solutions. There are two branches 
in both figures that correspond to two solutions of type II that we find. 
From Figs.~\ref{fig:resultsII}--\ref{fig:ENII}, we see that at
$\mu=0$ the minima of both branches (referred to as branches no.~1-2 below) is at $Q=0$ and have
\begin{eqnarray}
m^{(1)}_B/3&\simeq&1.0347, \quad {\rm with} \quad {\cal E}^{(1)}_{II}\simeq -1.6826,
\label{min1}\\
m^{(2)}_B/3&\simeq&0.1309, \quad {\rm with} \quad {\cal E}^{(2)}_{II}\simeq -0.8981.\label{min2}
\end{eqnarray}
Consequently, the 
ground state of Solution II at $\mu=0$ is given by the minimum of
branch no.~1 (\Eq{min1}).

To see whether the chiral wave instability takes place, we consider
Fig.~\ref{fig:ENII} and find that the minima
of ${\cal E}_{II}$ is at $Q=0$ for all values of $\mu$. This is not in
contradiction with the discussion
of Section~\ref{Seff_CDW} because, there, we only showed
that the baryon contribution to the
free energy, ${\cal E}_B(\Sigma,Q)$, has a minimum at $Q\neq0$. What we see
here is that the remaining contributions to ${\cal E}$
offset this minimum and lead to a ground state with $Q=0$.

It is important to note that even if the minimum of ${\cal  E}_B$ was deeper, such that ${\cal E}_{II}$ itself
had a minimum at $Q\neq 0$, it is still unlikely that the chiral
density wave would become the ground state. To understand
why, recall that in Section~\ref{Seff_CDW} we showed that what allows the crystalline phase to compete with the $\Sigma=0$ phase of
Solution I are the gapless excitations in the baryon spectrum. From \Eq{eq:Epm} we see that the
momentum of these excitations is given by 
\begin{eqnarray}
\vec p&=&(0,0,p_z),\\
\sin^2 p_z/2 &=& \sin^2 (NQ/2) - \Sigma^2,\label{cond}
\end{eqnarray}
which has a solution only if 
\begin{equation}
\Sigma \le 1.\label{cond1}
\end{equation}
Unfortunately, with the baryon mass of our ground state~(\ref{min1})
we have $\Sigma=\sinh m_B\sim 11$. This prohibits
gapless excitations, and makes the appearance of the chiral waves improbable.\footnote{The true relevant value of
  $\Sigma$ that should tested against \Eq{cond1} is
  $\Sigma(Q;\mu)\equiv \sinh m_B (Q;\mu)$, but from
  Fig.~\ref{fig:resultsII} we see that $\Sigma(Q;\mu)\ge
  \Sigma(\pi;0)\simeq 4$, for which \Eq{cond1} is still not satisfied.}

\subsubsection{Evolution of mean-field ground state as a function of $\mu$} 
\label{subsec:evolve}

Here we discuss the way the ground state evolves with increasing
chemical potential $\mu$. Given the conclusions of the previous
sections, we restrict to states with $Q=0$ only. 

At $\mu=0$, solution of type I has 
${\cal E}_I\simeq -1$, while the solution of type II that has the
lowest energy is given in \Eq{min1}. Comparing ${\cal E}_I$ and the energy in
\Eq{min1} we see that the latter is the ground state of
the system, and so, the baryon mass is  $m_B/3\simeq 1.0347$. The
condensates $V_q$ (\Eq{ansatz0}) that we find for this ground state 
 are 
\begin{eqnarray}
\<\bar \frac{\chi \chi}3\>&=&V_1\simeq 0.67685,\label{V1}\\
\<\left( \frac{\bar \chi \chi}3\right)^2\>&=&V_2\simeq 0.31947,\label{V2}\\
\<\left( \frac{\bar \chi \chi}3\right)^3\>&=&V_3\simeq 0.07872.\label{V3}
\end{eqnarray}
 Comparing these results to the literature, we see that 
our mass is $0.22\%-0.54\%$ lower than 
other tree level results 
(see, for example, the summary in
 \cite{PdF}), and that our condensates are close to the analytical and numerical results of
\cite{HKS,KS,Martin_Siu,Nishida,PdF} (discrepancies are on the level of
$2\%-5\%$).

\begin{figure}[htb]
\includegraphics[width=10cm]{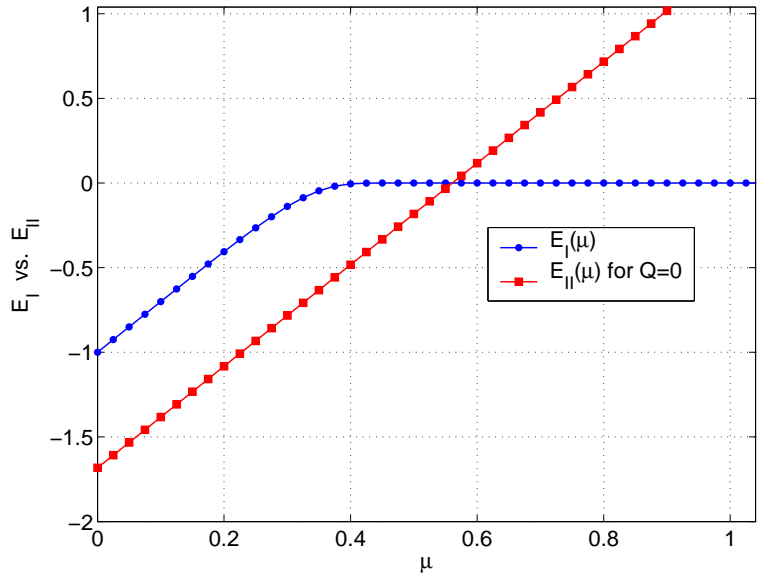}
\caption{The energy densities ${\cal E}_{I}$ and ${\cal E}_{II}$ as a
  function of $\mu$. 
\label{fig:EIvsEIIQ0}}
\end{figure}

When $\mu$ increases, the energy densities of both type
 of solutions grow. As explained in Section~\ref{subsec:SolI}, the
 energy of solution I stops increasing at $\mu_s$ where it reaches
 ${\cal E}_I=0$. Nothing special happens to solution II at
this point, and its energy is still
negative and continues to grow monotonically. In
Fig.~\ref{fig:EIvsEIIQ0} we replot ${\cal
  E}_{I}(\mu)$ of Fig.~\ref{fig:ENI} together with the ground state
 value of ${\cal
  E}_{II}(\mu)$ (i.e. for given $\mu$, we extract the minima of the
curves in Fig.~\ref{fig:ENII}).

From  Fig.~\ref{fig:EIvsEIIQ0} we see that a transition occurs at $\mu=\mu_t$ where 
\begin{equation}
{\cal E}_{II}(\mu_t)=0.\label{mu_t}
\end{equation}
Substituting\Eq{eq:EMF} and \Eq{eq:MFhV} in \Eq{mu_t} gives (for $Q=0$)
\begin{equation} 
3 (d+1)\sum_{q=1}^3 a_q\,
V_q^2= \int \left( \frac{dp}{2\pi}\right)^3 \left( 
  E-3\mu_t \right) \theta \left( 
  E-3\mu_t \right). \label{eq:determine_mu_t}
\end{equation} 
From Fig.~\ref{fig:resultsII} we see that, for $Q=0$, the mass $m_B$ does not change as a function
of $\mu$ up to $\mu\simeq m_B/3$, at which point solution II
disappears. This reflects the fact that the condensates $V_q$
 are independent of $\mu$ as well, and allows us to use the
$\mu=0$ values of $m_B$ and $V_q$ in 
\Eq{eq:determine_mu_t}. This is a 
consistent procedure provided that the
 resulting $\mu_t$ obeys
 $\mu_t<m_B/3$. 

Indeed, solving \Eq{eq:determine_mu_t} for $\mu_t$ gives
\begin{eqnarray} 
\mu_t &=& \left(\frac13 \int \left( 
  \frac{dp}{2\pi}\right)^3 E(p)\right) - 4 \left(a_1 V^2_1+a_2 V^2_2+a_3
  V^2_3\right) \nonumber \\
&\simeq& 1.0367-\left(0.45813 + 0.01914 -0.001452 \right)= 0.5609, \label{eq:mu_t}
\end{eqnarray} 
which is lower than $m_B/3\simeq 1.0347$. Consequently this means that
the chiral symmetry of \Eq{U1eps} is restored in a first order
transition at $\mu_t\simeq 0.5609$ that separates a low density phase characterised by the baryon
mass of \Eq{min1} and the chiral condensates in
Eqs.~(\ref{V1})-(\ref{V3}), and a high density phase where this
symmetry is intact and the baryons are massless.

Let us emphasise that the way chiral symmetry is restored 
here is largely influenced by lattice artifacts of the infinite
coupling limit. In particular, we cannot
observe the crystalline phase partly because the mass of the baryons
 in lattice units, $m_B$, is around $\sim 3$. This prevents the
 baryons spectrum from having gapless excitations,
and makes their contribution to the free energy ${\cal E}_B(m_B,Q)$ higher compared to what
it would be with $m_B=Q=0$. This is in contrast to the
`continuum-like' scenario that we discussed in Section~\ref{Seff_CDW},
where $m_B$ was small, and ${\cal E}_B(m_B,Q)<{\cal E}_B(0,0)$. 
 In the next section we study such a continuum-like case by
 taking a formal limit of $S_{\rm HKS}$, in which the mass of the $b$ fields is significantly smaller than 
discussed above.

\subsection{The formal limit of $N\to1$}
\label{sec:PD_N1}

In this section we study a formal limit of \Eq{eq:Seff}, where one takes
$N=1$ and $F_N(u)= u/4$. In this case \Eq{eq:Seff} cannot be obtained from an
underlying lattice gauge theory, but nevertheless seems to be useful to
understand, since it shows chiral 
restoration with less lattice artifacts. In their original preprint, the authors of \cite{HKS} have
considered this limit as an example for a case where
the baryonic terms in \Eq{eq:Seff} are important at $\mu=0$. 
Here we approach this limit at nonzero $\mu$ and seek for signs of a
crystalline solution for the mean-field equations. We first do so for
$d=1$, where it is known that crystalline instabilities are robust \cite{Peierls,Overhauser}. (The
physical significance of the result in this case is, however, 
unclear since the continuous $U(1)_\epsilon$ symmetry cannot break in $1+1$ dimensions.)

We begin the analysis by noting that for $N=1$, \Eq{eq:Seff}
depends only linearly on $m_n m_{n+\hat \nu}$. This allows us to take 
$h_2=h_3=V_2=V_3=0$, and to call $h_1=h,V_1=V$. 
The mean-field equations become (we keep $d$ general at this stage)
\begin{eqnarray} 
h&=&2(d+1)a_1V\gamma_1(Q), \label{eq:MFeq1_N1} \\
V&=&A(h,Q;\mu), \label{eq:MFeq2_N1}
\end{eqnarray} 
and the free energy is now given by
\begin{eqnarray} 
{\cal E}= \frac{h^2}{4(d+1)a_1\gamma_1(Q)}+ {\cal 
E}_{\text{matter}}. \label{eq:EMF_N1}
\end{eqnarray}
Both $A(h,Q;\mu)$ and ${\cal E}_{\rm matter}$ are still defined in
Eqs.~(\ref{eq:A}), and (\ref{eq:det}), with the difference that the
dependence of $\Sigma_n$ on $h_n$ is now
\begin{equation}
\Sigma_n=h_n=h e^{-i\epsilon_n  Q \, n_1}.
\end{equation}

\begin{figure}[htb]
\includegraphics[width=13cm]{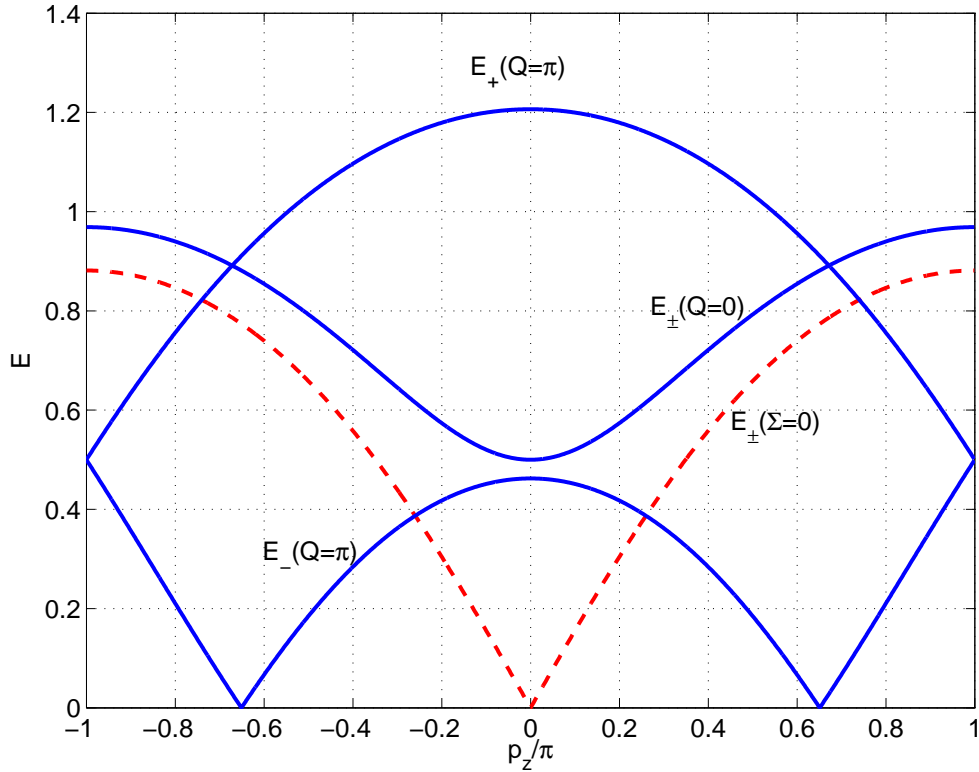}
\caption{The energy bands $E_\pm$ of the baryons for
  $p=(0,0,p_z)$. The dashed line is
  the gapless band with $m_B=0$, and
 the solid lines are have $m_B=0.5$, and $Q=0,\pi$.
\label{fig:Epm_N1}}
\end{figure}

Again, it is easy to see that there are two type of solutions which
are the analogues of the type I and II solutions in the $SU(3)$ case. In Fig.~\ref{fig:Epm_N1} we
present the energy bands for $m_B=\sinh^{-1}(\Sigma)=0.5$, which is close to the value that solves the mean-field equations for $N=1$ and $d=1$.
 With this value of $m_B$ the effect of the lattice coarseness is
smaller compared to the $N=3$ case 
: the massive and massless bands of solutions I and II overlap, and
there exists momenta where the energy of solution II is zero.

\begin{figure}[htb]
\includegraphics[width=12cm]{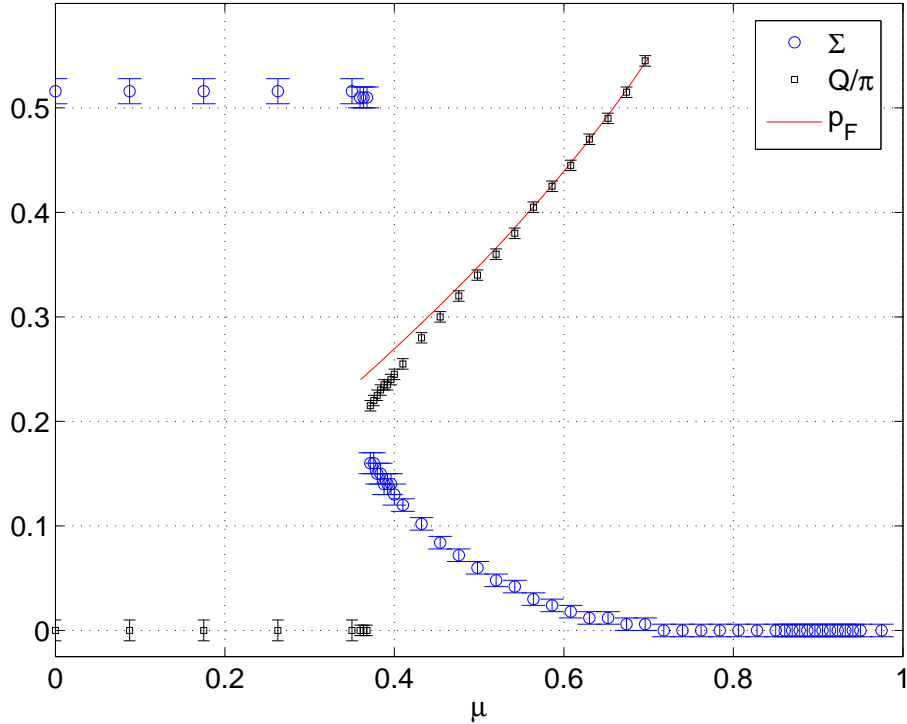}
\caption{The values of $V$ (circles), and $Q/\pi$ (dots) that solve
  the mean-field equations~(\ref{eq:MFeq1_N1})--(\ref{eq:MFeq2_N1}) in
  $1+1$ dimensions. The red solid line
 denotes the analytic expression of \Eq{Q_mu}. Note that we present $Q/\pi$ (black square symbols)
only when $\Sigma>0$, since otherwise $S_I^{\rm mean-field}=0$, and $Q$ is ill-defined.
\label{fig:resultsII_N1}}
\end{figure}

As a result, a crystalline ground state is stable. 
In Fig.~\ref{fig:resultsII_N1} we present the dependence of $V$ and
$Q$ on $\mu$. 
We also plot the following analytic expectation for $Q(\mu)$ that
arises from our
discussion in Section~\ref{Seff_CDW} (adjusted to $N=1$)
\begin{equation}
Q(\mu)=p_F(\mu)=2\, \sin^{-1} \left(\sinh \mu\right). \label{Q_mu}
\end{equation}
It is clear
that \Eq{Q_mu} works well, especially for $\Sigma\ll 1$, which means that the mechanism
described in Section~\ref{Seff_CDW} is indeed the one
generating the chiral density wave instability here. 

When we move from $d=1$ to $d=2$ the crystalline phase
disappears. We find that the reason is that the meson self energy described by
the first term in \Eq{eq:EMF_N1} prefers $Q=0$, and `wins' the
instability generated in ${\cal E}_{\rm
matter}$. This effect comes from the function
$\gamma_1(Q)$ in \Eq{gammaQ}, and can be considered as a lattice artifact as well,
since in the continuum limit the lattice quantities $\Sigma, \mu$, and $Q$ vanish, and 
so $\gamma_1(Q)\to 1$. Indeed, in continuum treatments of the Nambu-Jona-Lasinio
model such as \cite{NJL_paper}, the contribution to the free energy that is the
analog of the first term in \Eq{eq:EMF_N1}, does not depend on $Q$.

When we move to $d=3$ it appears that even without the $Q$ dependence
in $\gamma_1(Q)$, the crystalline phase is unstable. This changes if one makes the
meson self-interaction $F_1(u)=\frac14 u$ stronger by changing it to
$F_1(u)=a_1 u$ with $a_1>1/4$.

\section{Conclusions}
\label{sec:summary}

Our aim in this paper was to see whether the phase diagram of
strongly-coupled lattice QCD at zero temperature and nonzero density
includes an incommensurate crystalline phase with chiral density
waves. 
To do so we followed Hoek,
Kawamoto, and Smit (HKS), and used the effective hadronic action, $S_{\rm HKS}$,
that they derived in \cite{HKS} for one-component staggered
fermions. We formulated
a mean-field theory for $S_{\rm HKS}$ which is novel
in the following ways. Firstly, we do not neglect the terms in $S_{\rm HKS}$ that have 
four and six powers of the meson fields, and  treat the baryon
contribution in full. Secondly, our mean-field analysis is free of any
undetermined parameters and we fix all vacuum expectation values by
minimising the free-energy. Lastly, we do not assume a homogeneous
ansatz, which is imperative in order to look for the crystalline
phase. This leads us to introduce auxiliary fields without the usual
Hubbard-Stratonovich transformation, that can become ill-defined for
inhomogeneous vacua. 

Despite the fact that the generic structure of $S_{\rm HKS}$ can give
rise to a crystal phase (see discussion in
Section~\ref{Seff_CDW}), we find that this {\em does not} happen with
our mean-field ansatz. This is partly due to lattice
artifacts that come in the form of a large lattice baryon mass $m_B\sim
3$. The latter prohibits any
possibility to have a gapless baryonic spectrum in the presence of the
chiral density waves, which would allow the
crystalline vacuum to compete with the massless homogeneous vacuum. To check this
explanation, we take a formal limit of $S_{\rm HKS}$, where $m_B\simeq
0.5$, and indeed find that this lower mass can stabilise the crystal. 

More precisely, when we study this formal limit in the $1+1$ system, we see a crystal
structure at intermediate densities, with a wave vector given by the Fermi momentum of the system. However, this result is
not robust; when we move to $2+1$ and $3+1$ dimensions we see that the
crystalline instability is too weak to survive. The reasons for
that include an additional lattice artifact, that does not go away when $m_B$ decreases, and a too weak interaction between the fermions.

Our results stress that to get continuum physics from
Monte-Carlo simulations at nonzero $\mu$ and low temperature, one will
have to simulate on relatively fine lattices, with weak couplings. 
In particular, the baryon mass in
lattice units, $m_B$, should be much smaller than its strong-coupling limit of
$m_B\sim 3$. Only when this happens will the structure of the baryon
energy bands be `continuum-like'. For stronger
couplings, the energy bands are {\em qualitatively} different than the continuum ones, and 
physical phenomena that are sensitive to their structure will also be {\em
  qualitatively} different than it is in the continuum. 
Chiral crystals is an
good  example for such a phenomenon, 
that simply goes away at strong
couplings.

There are several ways in which one could 
extend this study. First, recall that we have 
focused on an 
ansatz that breaks translation invariance only in one direction. This direction was chosen to be
along one of the lattice axes, but a further study can generalise this
easily (we have already derived the free energy for a general $\vec
Q=(Q_1,Q_2,Q_3)$) and check whether this
 makes the crystalline phase more robust. Second, as the authors in \cite{Shuster}
point out, one can lower the energy of the crystalline
phase by considering an ansatz which is a linear combination of
waves, that will break translation invariance in all directions. 
Third, it will be
useful (but also hard) to see how do the sub-leading terms in the
strong-coupling expansion influence the results. These were obtained in
\cite{Ichinose} for the fermions we consider here, and give rise to
more complicated interactions between the hadrons. Finally, it can be 
interesting to numerically study the possibility of chiral density waves in the
Nambu-Jona-Lasinio model, where Monte-Carlo simulations are free from
the sign-problem.

\begin{acknowledgments} 
I thank J.~Burkardt for help 
on using his numerical integration
 routines,  J.~Chalker and F.~Essler 
for useful discussions, and Y.~Shamir and 
B.~Svetitsky for their remarks on this manuscript.
I was supported by PPARC. 
\end{acknowledgments}

\appendix

\section{The baryonic determinant} 
\label{sec:baryon_det}

In this section we calculate the baryonic determinant \Eq{eq:det} and its
variation $A(h_q,Q;\mu)$ of \Eq{eq:A}. We begin
by showing that $A$, which, for the general $SU(N)$ case, is defined by
\begin{eqnarray}
A(h_q,Q;\mu)&\equiv& \<\bar{b}_n b_n \> \,e^{-Ni\vec Q \vec n\, \epsilon_n},\\ \label{eq:A1}
\<\bar{b}_n b_n \>&=&\frac{\int D\bar b D b \,\, e^{S_0} \, \, \bar
  b_n b_n}{\int D\bar b D b \,\, e^{S_0}} \\
S_0&=&\sum_{n,m} \bar b_n \left[ \Sigma_n \bm{1} + D_{nm} \right] b_m,
\end{eqnarray}
is independent of the lattice site index $n$. To see this we use the symmetries of the
action $S_0$
\begin{equation} 
S_0\equiv \bar{b} \, K \, b = \sum_n \bar{b}_n\, \Sigma e^{-iN \vec Q\vec n\epsilon_n} \,b_n + \frac12 
\sum_{n\nu} \bar{b}_n \left( \eta_{n\nu}   b_{n+\hat \nu} 
- \eta^{-1}_{n\nu} b_{n-\hat \nu} \right) \equiv S_\Sigma + S_{\rm kinetic},
\end{equation} 
where $\epsilon_n=\pm$ is the parity of the site. In particular,
$S_{\rm kinetic}$ is invariant
under spatial translations
\begin{equation}
T_{\vec R} \quad : \quad  n\to n+\vec R
\end{equation}
{\em and} chiral symmetry 
\begin{equation}
U(1)_\epsilon \quad : \quad b_n\to
e^{iN\theta \epsilon_n}b_n,\quad \bar{b}_n\to
\bar{b}_n e^{iN\theta \epsilon_n}.
\end{equation}
The term $S_\Sigma$ is invariant under these
symmetries only for the correlated transformation with $2\theta=\vec Q\cdot
 \vec R$. (More precisely, if $R$ takes one from an even site to an odd site
($\epsilon_n=-\epsilon_{n+R}$) then an extra transformation of
reflection around the origin is needed to keep
$S_\Sigma$ unchanged.)
 
Although $S_0$ is invariant under these correlated transformations,
$\< \bar{b}_n b_n\>$ is not, and transforms as follows
\begin{eqnarray}
{\text For}\quad  \epsilon_n&=&\epsilon_{n+\vec R} \quad : \quad \<\bar{b}_n b_n\>
\stackrel{\rm T_{\vec R}}
     {\longrightarrow}
\<\bar{b}_{n+\vec R} b_{n+\vec R}\> 
\stackrel{ U(1)_\epsilon}
     {\longrightarrow}
\<\bar{b}_{n+\vec R}\,
b_{n+\vec R}\> \, e^{-Ni\vec Q\vec R\epsilon_n} \label{eq:tranf1} \\
{\text For} \quad \epsilon_n&=&-\epsilon_{n+\vec R} \quad : \quad  \<\bar{b}_n
b_n\> 
\stackrel{\rm T_{\vec R}}
     {\longrightarrow}
\<\bar{b}_{n+\vec R} b_{n+\vec R}\>
\stackrel{ U(1)_\epsilon}
     {\longrightarrow}
\<\bar{b}_{n+\vec R}\,
b_{n+\vec R}\> \, e^{-Ni\vec Q\vec R\epsilon_n} \nonumber \\
&&
\stackrel{\rm reflection}
     {\longrightarrow}
\<\bar{b}_{-(n+\vec R)}\, b_{-(n+\vec R)}\> \, e^{-Ni\vec Q\vec R\epsilon_n} = 
\<\bar{b}_{-(n+\vec R)}\, b_{-(n+\vec R)}\> \, e^{+Ni\vec Q\vec R\epsilon_{-(n+\vec R)}}.\label{eq:tranf2}
\end{eqnarray}

Putting $n=0$ gives $\<\bar{b}_0 b_0\> = \<\bar{b}_{\vec R} b_{\vec R}\>\,e^{-Ni\vec Q\vec R\epsilon_{\vec R}}$ for $\epsilon_{\vec R}=+1$ from \Eq{eq:tranf1}, and for $\epsilon_{\vec R}=-1$ from \Eq{eq:tranf2}. This proves that
\begin{equation}
\<\bar{b}_n b_n \> \, e^{-N i\vec Q \vec n \epsilon_n } = {\text independent \,\,\, of \,\,\, n},
\end{equation}
and therefore that\footnote{This is different from \Eq{eq:A} since
  here we evaluate the determinant for the helical ansatz, and
 then take the derivative with respect to $\Sigma$.}
\begin{equation}
A(h_q,Q;\mu)= \frac1{N_s} \frac{\partial \log \det \left[ \Sigma \,
    e^{-Ni\vec Q\vec n\epsilon_n }
    \bm{1} + D\right] }{\partial \Sigma}.
\end{equation}

We now turn to calculate the determinant itself. It will be convenient to
note that 
\begin{equation}
\log \det K\equiv \log \det \left[ \Sigma e^{-iN\vec Q\vec
    n\epsilon_n} \bm{1} + D \right] =
\log \det \left[ \bm{1} + \Sigma^{-1} e^{Ni\vec Q\vec n\epsilon_n}  D \right] + \sum_n \log
\Sigma - Ni \sum_n \vec Q\vec n \epsilon_n, \label{eq:K2Ktilde}
\end{equation}
and to calculate $\det \tilde{K}\equiv \det \left[ \bm{1} +
  \Sigma^{-1} e^{Ni\vec Q\vec n\epsilon_n}D \right]$
instead. (Note that the last term in \Eq{eq:K2Ktilde} drops out for a
lattice with an even number of sites in each direction). To proceed,
we use coordinates defined on a new lattice with spacing $a=2$, and write 
$n=2X+\rho$, with $\rho_{1,\dots,d}=0$ or $1$ and $X$ taking 
values in the new lattice \cite{Rothe}. In these coordinates we use 
the following definitions 
\begin{eqnarray} 
b_n&\equiv& b_\rho(X),\\ 
b_{n+\hat \nu}&\equiv& \sum_{\rho'}\left(\delta_{\rho',\rho+\hat \nu} b_{\rho'}(X) + \delta_{\rho',\rho-\hat \nu} 
b_{\rho'}(X+\hat \nu) \right),\\ 
b_{n-\hat \nu}&\equiv& \sum_{\rho'}\left(\delta_{\rho',\rho-\hat \nu} 
b_{\rho'}(X) + \delta_{\rho',\rho+\hat \nu} b_{\rho'}(X-\hat \nu) \right). 
\end{eqnarray} 
Moving to momentum space with 
\begin{eqnarray} 
b_\rho(X)&=&\sqrt{\frac{2^d}{N_s}} \sum_{p} e^{i\left(pX + \epsilon_\rho 
\vec Q(2\vec X+\vec \rho)/2\right)} b_\rho(p),\\
\bar b_\rho(X)&=&\sqrt{\frac{2^d}{N_s}} \sum_{p} e^{-i\left(pX + \epsilon_\rho 
\vec Q(2\vec X+\vec \rho)/2\right)} \bar b_\rho(p),
\end{eqnarray} 
we get
\begin{eqnarray} 
\tilde{S}_0 &\equiv& \bar{b} \, \tilde{K} \, b = 
\sum_{\rho,\rho'\atop p} \bar{b}_\rho(p) \tilde K_{\rho\rho'}(p) 
b_\rho(p),\\ 
\tilde K_{\rho\rho'}(p) &=& \delta_{\rho\rho'}+\Sigma^{-1}\sum_\nu \left(\eta_\nu e^{ip'\nu/2} -
    \eta^{-1}_\nu e^{-ip'\nu/2} \right)\left(\delta_{\rho+\hat \nu,\rho'} +
    \delta_{\rho-\hat \nu,\rho'} \right)\e^{ip(\rho-\rho')/2},\label{eq:K_tilde1} \\
p'&=&p-NQ\epsilon_\rho,
\end{eqnarray} 
where we used $\epsilon_\rho=-\epsilon_{\rho'}$. \Eq{eq:K_tilde1} can be brought to the form
\begin{eqnarray} 
\tilde K(p) &=& \bm{1} + i\Sigma^{-1} \sum_{\nu=0}^{d} \Gamma_\nu(p) \, \sin 
\left(p_\nu/2 - i N\mu \delta_{\nu,0} + NQ_\nu 
\hat \epsilon/2 \right)\\ 
&=&\bm{1} + i\Sigma^{-1} \left[ \Gamma_0(\omega) \sin \left( 
\omega/2-iN\mu \right) \right.\nonumber\\
&&+\left. \sum_{\nu=1}^{d} \Gamma_\nu(p) \, \sin (p_\nu/2)\cos
  (NQ_\nu/2) +  \Gamma_\nu(p)\,\, \hat \epsilon \, \cos (p_\nu/2) \sin (NQ_\nu/2) \right] 
\end{eqnarray} 
Here we defined 
\begin{equation} 
\left(\hat \epsilon\right)_{\rho\rho'} \equiv \epsilon_\rho \delta_{\rho\rho'}, 
\end{equation} 
and also 
\begin{equation} 
\left(\Gamma_\nu\right)_{\rho\rho'} = \tilde \eta_\nu(\rho) \left( 
\delta_{\rho,\rho'+\hat \nu} + \delta_{\rho,\rho'-\hat \nu} \right) 
e^{ip(\rho-\rho')/2}, 
\end{equation} 
where $\tilde \eta_\nu=\eta_\nu$ for $\nu\in [1,d]$ and $\tilde \eta_0=1$. 
The matrices $\Gamma_\nu$ obey
\begin{equation} 
\left\{ \Gamma_\nu,\Gamma_\tau \right\} = 2 \delta_{\nu\tau} 
\bm{1}, \qquad \left\{ \Gamma_\nu,\hat \epsilon \right\}= \left\{ \Gamma_\nu,\Gamma_{5\tau} \right\} = 0, 
\end{equation} 
with 
\begin{equation} 
\left(\Gamma_{5\nu}\right)_{\rho\rho'} = i\tilde \eta_\nu(\rho) \left( 
\delta_{\rho,\rho'+\hat \nu} - \delta_{\rho,\rho'-\hat \nu} \right) 
e^{ip(\rho-\rho')/2}, 
\end{equation} 
that also obey 
\begin{equation} 
\left\{ \Gamma_{5\nu},\Gamma_{5\tau} \right\} = 2 \delta_{\nu\tau} 
\bm{1}, \qquad \left\{ \Gamma_{5\nu},\hat \epsilon \right\} = 0. 
\end{equation} 

 Next we use the fact that $\det \hat \epsilon \, \tilde K(p) \, \hat \epsilon 
\, \tilde K(p)=\det \tilde K^2(p)$ and that $\left\{\Gamma_\nu(p),\hat
  \epsilon
\right\}=0$ to get 
\begin{equation} 
\left( \det \tilde K(p) \right)^2 = \det \hat \epsilon \, \tilde K(p)\, 
\hat \epsilon  \times 
\det \tilde K(p) \equiv \det K_2(p), 
\end{equation} 
with 
\begin{equation} 
K_2(p)=\left(\epsilon \, \tilde K(p)\, \hat \epsilon\right)  \tilde K(p) =\left[\bm{1}\left( 1 + \Sigma^{-2} \left(\sum_{\nu=0}^{d} 
      s^2_\nu -\sum_{\nu=1}^d c^2_\nu \right) \right) + 2\Sigma^{-2} \sum_{\mu\neq
    \nu\atop \mu\neq 0} s_\nu c_\mu \Gamma_\nu \Gamma_\mu \,\hat \epsilon \right].
\end{equation} 
Here we defined
\begin{eqnarray}
s_0&=&\sin \left(p_0/2 -iN\mu\right),
\end{eqnarray} 
and for $\nu>0$ 
\begin{eqnarray}
s_\nu&=&\sin(p_\nu/2)\cos (NQ_\nu/2),\\
c_\nu&=&\cos(p_\nu/2)\sin (NQ_\nu/2).
\end{eqnarray} 
 To proceed we need to find a matrix that 
anti-commutes with $\Gamma_\nu \Gamma_\mu \hat \epsilon$ for all 
$\nu\neq\mu$. We can choose any of the matrices 
$M=\Gamma_{5\nu}$ (which also obey $M^2=\bm{1}$) and we have
\begin{equation} 
\left( \det \tilde K(p)\right)^4=\left( \det K_2(p) \right)^2=\det M 
K_2(p) M \times \det K_2(p)\equiv \det K_4(p), 
\end{equation} 
with 
\begin{eqnarray} 
K_4(p)&=&\left(MK_2M\right)K=\left[\bm{1}\cdot \left( 1 + \Sigma^{-2} \left(\sum_{\nu=0}^{d} 
      s^2_\nu -\sum_{\nu=1}^d c^2_\nu \right) \right)^2-4\Sigma^{-4} C^2\right],\nonumber\\
C&=&\sum_{\mu\neq
    \nu\atop \mu\neq 0} s_\nu c_\mu \Gamma_\nu \Gamma_\mu \,\hat \epsilon.
\end{eqnarray} 
Using the anti-commutation relations of $\Gamma_\nu$ we get $C^2=-B\cdot \bm{1}$ with
\begin{equation}
B=s^2_0\sum_{\nu=1}^d c^2_\nu + \sum_{\nu\neq\mu\atop \nu,\mu\ge 1}\left(s_\nu c_\mu-c_\nu s_\mu \right)^2,
\end{equation}
and defining $D(p)\equiv \det K_4(p)$ we obtain the following form
\begin{eqnarray} 
D(p)\cdot \Sigma^4&=&\left(\sin^2(p_0/2-iN\mu)+\epsilon^2_+(p)
\right) 
\times 
\left(\sin^2(p_0/2-iN\mu)+ \epsilon^2_-(p) \right).
\end{eqnarray} 
Here
\begin{eqnarray}
\epsilon^2_\pm&=&|\vec{s}|^2\sin^2 \phi + \left(
  \sqrt{\Sigma^2+|\vec{s}|^2\cos^2 \phi} \pm 
|\vec{c}|\right)^2, \\
s_i&\equiv&\sin (p_i/2)\cos (NQ_i/2),\\
c_i&\equiv& \cos (p_i/2)\sin (NQ_i/2),\\
\cos \phi&\equiv&\hat{c}\cdot \hat{s}.
\end{eqnarray} 

Since $\det \tilde K(-p)=\det \tilde K^*(p)$ then 
 $\det \tilde K
\equiv \prod_p \det \tilde K(p)$ is real\footnote{This can be proved as 
follows. We have $\tilde K(p,Q)=\tilde K^*(-p,-Q)$, but also 
$\det \tilde K(p,Q)=\det \Gamma_{5\mu} \hat \epsilon \tilde K(p,Q)
\hat \epsilon \Gamma_{5\mu}=\det \tilde K(p,-Q)$. This means that 
$\det \tilde K(p,Q)\det \tilde K(-p,Q)=\det \tilde K(p,Q)\det \tilde K^*(p,-Q)=\det \tilde K(p,Q)\det \tilde K^*(p,Q)=|\det \tilde K(p,Q)|^2$ is real.} 
 and we can 
write $\det \tilde K = \left( \det \tilde K^4 \right)^{1/4}=\left( \prod_p \det 
K_4(p) \right)^{1/4}$ or 
\begin{eqnarray} 
\log \det \tilde K &=& \frac14 \sum_p \log (D(p))^{2^d}=\frac{N_s}4 
\int_{-\pi}^{+\pi} \left( \frac{dp}{2\pi} \right)^{d+1} \log D(p). 
\end{eqnarray} 
  
Dropping irrelevant constants, we get 
\begin{eqnarray} 
\frac1{N_s}\log \det K &=& \frac{1}{4} \sum_{b=\pm} 
\int_{-\pi}^{+\pi} \left( \frac{d\vec{p}}{2\pi}
 \right)^{d} 
\int_{-\pi}^\pi \frac{dp_0}{2\pi}
 \log 
\left[\sin^2(p_0/2-iN\mu)+\epsilon^2_b
\right],\nonumber \\ \label{eq:logdetK} \\ 
 \frac1{N_s}\frac{\partial \log \det K}{\partial \Sigma} &=&
 \frac{\Sigma}{2} 
\sum_{b=\pm} \int_{-\pi}^{+\pi} \left( 
\frac{d\vec{p}}{2\pi} \right)^{d} \int_{-\pi}^\pi 
\frac{dp_0}{2\pi} \frac{\left(1+b\frac{|\vec
      c|}{\sqrt{\Sigma^2+\vec s^2\cos^2 \phi}} 
 \right)}{\sin^2(p_0/2-iN\mu)+\epsilon^2_b(p)}. \label{eq:d_logdetK}
\end{eqnarray} 

Performing the change of variables $z\equiv e^{ip_0 -2N\mu}$ 
the integral in \Eq{eq:d_logdetK} becomes a simple contour integral
 and using complex analysis one can show that
\begin{eqnarray} 
\int_{-\pi}^\pi \frac{dp_0}{2\pi}\, 
\frac{1}{\sin^2(p_0/2-iN\mu)+
\epsilon^2_b}&=&\frac{\theta(E_b(p)-N\mu)}{\sqrt{\epsilon^2_b(1+\epsilon^2_b)}},
\end{eqnarray}
 where $\sinh^2 E_b(p,s) = \epsilon^2_b(p)$.
The integration in 
\Eq{eq:logdetK} is carried out as follows (dropping an irrelevant constant of integration)
\begin{eqnarray}
\int_{-\pi}^\pi \frac{dp_0}{2\pi}\, \log 
\left[\sin^2(p_0/2-iN\mu)+\epsilon^2_b(p)\right]&=&\int_{X_0}^{\epsilon_b}  dX \int_{-\pi}^\pi \frac{dp_0}{2\pi}\, 
\frac{1}{\sin^2(p_0/2-iN\mu)+X} \nonumber \\
&=&2\left[ E_b(p)-N\mu \right] \theta( E_b(p)-N\mu).
\end{eqnarray} 
 The final result is then 
\begin{eqnarray} 
\frac1{N_s}\frac{\partial \log \det K}
{\partial \Sigma}&=&\frac{\Sigma}2\sum_{b=\pm} \int 
\left(\frac{dp}{2\pi}\right)^d 
\frac{\left(1+b\frac{|\vec c|}{\sqrt{\Sigma^2+|\vec s|^2\cos^2 \phi}}\right)}{\sqrt{\epsilon^2_b(1+\epsilon^2_b)}} 
\,\, \theta \left( E_b(p)-N\mu\right)\nonumber \\ \\ 
\frac1{N_s}\log \det K &=&\frac12 \sum_{b=\pm} \int \left(\frac{dp}{2\pi}\right)^d \left(E_b(p)-N\mu\right) 
\theta \left( E_b(p)-N\mu\right). \nonumber 
\\ 
\end{eqnarray} 
which we evaluate numerically. In particular, in the case of $d=3$,
 we present results for $\vec Q||\hat{z}$, in which case 
\begin{eqnarray}
\epsilon^2_\pm&=&|\vec{s}_\perp|^2 + \left(
  \sqrt{\Sigma^2+\vec{s}_z^2} \pm 
|c|\right)^2, \\
s_z&=&\sin (p_z/2)\cos (NQ/2),\\
c&=& \cos (p_z/2)\sin (NQ/2),\\
|\vec{s}_\perp|^2&=&\sin^2 p_x/2 + \sin^2 p_y/2.
\end{eqnarray}  
Since the transverse directions appear very simply here, we define the
density of states
$\displaystyle{D(s)ds = 
\left(\frac{dp}{2\pi}\right)^{d-1} \delta 
\left(\sum_{\nu=x,y}\sin^2 (p_\nu/2) -s\right)}$ for the transverse
momenta. One can show that $D(s)=\frac2{\pi^2}K(s(2-s))$ where $K(x)$
is the complete elliptic integral of the first kind (for example see \cite{DOS}).

\end{document}